\documentclass[11pt,a4paper]{article} 

\usepackage{natbib}
\usepackage{parskip} 
\usepackage{titlesec}
\usepackage{stmaryrd}
\usepackage{graphicx}
\usepackage{caption}
\usepackage{subcaption}
\usepackage{amssymb}
\usepackage{enumitem}
\usepackage{bbm}
\usepackage{bm}
\usepackage{verbatim}
\usepackage{enumerate}
\usepackage[hyperindex]{hyperref}
\usepackage{amsmath, amsfonts, amsthm, amscd}       
\usepackage{fancybox}
\usepackage{xcolor}
\usepackage{fix-cm}
\usepackage[T1]{fontenc}
\usepackage[ruled,vlined]{algorithm2e}
\usepackage{subcaption}
\newlength{\querylen}
\theoremstyle{plain}

\newtheorem{rmk}{Remark}

\newtheorem{?}{Question}

\usepackage{enumitem}

\usepackage{color}

\usepackage{titlesec}
\titleformat{\subsection}[block] 
    {\normalfont\bfseries\filright}
    {\thesubsection.}
    {0.5em}
    {}
\titlespacing*{\subsection}
    {0pt}
    {3.25ex plus 1ex minus .2ex}
    {3ex plus .5ex} 

\makeatletter

\usepackage{setspace}

\makeatother

\begin{document}
\title{Bayesian Posterior Learning of Mixed Graphical Models}
\author{Erdong Guo\thanks{Department of Statistical Science, University College London, Email: 
erdong.guo.22@ucl.ac.uk}, 
Alexandros Beskos\thanks{Department of Statistical Science, University College London}, and Maria De Iorio\thanks{Yong Loo Lin School of Medicine, National University of Singapore}}
\date{} 
\maketitle
\begin{abstract}
Mixed Graphical Models (MGMs) provide a flexible framework for structure learning from heterogeneous data by treating sets of both continuous and discrete.  
Bayesian inference for MGMs remains challenging due to the combinatorial complexity of graph space exploration and posterior computation.
In this paper, we model discrete components through latent Gaussian variables and consider two likelihood specifications: 
a copula-based ranked likelihood, yielding the copula-MGM, and a probit formulation based on cut-off points, yielding the probit-MGM. 
We then propose the \emph{Mixed Graph WWA}, a class of MCMC methods for posterior simulation in Bayesian MGMs. 
Building upon the WWA algorithm, we develop two specialized algorithms: copula-WWA for copula-MGMs and probit-WWA for probit-MGMs.
Both methods exploit the latent Gaussian representations to perform posterior inference through a Gibbs sampling scheme that alternates between latent-variable augmentation and graph-structure updates.
Through extensive simulation studies we demonstrate that the proposed methods achieve graph recovery accuracy comparable to or better than existing approaches, 
including copula-BD MCMC and probit-BD MCMC based on the Birth-Death MCMC methodology, 
while maintaining efficient posterior exploration and favorable effective sample size per unit computational time.
We further illustrate the practical utility of our approach through an application to the PAM$50$ breast cancer gene expression dataset, 
where the inferred MGMs reveal meaningful dependencies between gene expression profiles and cancer subtypes. 
These results highlight the effectiveness of the Mixed Graph WWA method as a scalable and principled tool for Bayesian structure learning in MGMs.
\end{abstract}

\section{Introduction}
Graphical models (GMs) provide a powerful framework for representing the dependence structure of high-dimensional multivariate data~\citep{lauritzen1996graphical}.
For a GM, nodes correspond to random variables, 
while edges encode the conditional dependencies between them, 
facilitating the abstraction of complex statistical relationships within high-dimensional datasets. 
Focusing on undirected GMs, two primary model classes emerge for homogeneous variates:
Gaussian Graphical Models (GGMs) for multivariate normal data~\citep{dempster1972covariance},
and discrete log-linear models for discrete variates~\citep{hofling2009estimation, ravikumar2010high}.
In GGMs, the conditional independence structure is fully characterised by the precision matrix,
whereas in discrete log-linear models dependencies are captured through interaction terms within an exponential family representation.
In both cases, presence of a zero entry in the precision matrix or absence of an interaction term implies a conditional independence relationship between the corresponding nodes.

Despite the wide applicability of the above models,
real-world datasets often contain \emph{both} continuous and discrete variables, necessitating the development of Mixed Graphical Models (MGMs) that can accommodate heterogeneous data types.

\begin{rmk}
The term `discrete' variables is used in this work to represent (standard) discrete, ordinal or binary variables. 
\end{rmk}

The foundation of MGMs can be traced back to the Conditional Gaussian Graphical Model (CGGM)~\citep{lauritzen1989mixed},
which introduced a principled approach of integrating both continuous and discrete nodes within a unified graphical framework.
A CGGM is formulated through a  conditional Gaussian density for the continuous variates given the discrete ones, with the conditional dependencies between variables governed by canonical parameters that adhere to the structural constraints imposed by the underlying graph.
Despite the theoretical elegance of the model,
a major limitation of the CGGM is that the number of parameters can grow exponentially fast with the number of variables,
making its direct application to real-world datasets computationally impractical.

A series of methodological advancements have been proposed to adapt the CGGM while maintaining its ability to capture the underlying structure of MGMs.
One prominent direction in this line of research is the simplification of the original CGGM to improve scalability while preserving flexibility.
\cite{lee2015learning, fellinghauer2013stable} developed MGMs under the pairwise Markov Random Field (MRF) assumption,
thereby reducing model complexity without significantly compromising structural expressiveness.
A key simplification introduced in subsequent works~\citep{cheng2017high} involves two main modifications:
(i) restricting interactions between binary variables to pairwise relationships and eliminating higher-order dependencies, 
and (ii) modeling the conditional covariance matrix and the canonical mean vector of the Gaussian variables as linear functions of the binary variables, rather than allowing arbitrary dependence.
Another line of research extends MGMs by modeling the conditional distributions using univariate exponential family distributions,
where the parameters are governed by the underlying graphical structure~\citep{yang2014mixed, chen2015selection}.
This approach leverages the flexibility of the exponential family to accommodate various data types while maintaining  interpretability of the GM.

In contrast to the aforementioned methods, an alternative class of MGMs employs hierarchical structures to obtain discrete variates via transforms of continuous ones.
Within this framework, MGMs with Gaussian latent variables have been widely explored, primarily through two modeling strategies:
(i) methods based on Gaussian copulas and use of Hoff's extended ranked likelihood~\citep{dobra2011copula, mohammadi2017bayesian},
and (ii) probit-based approaches~\citep{guo2015graphical, suggala2017ordinal}.
The copula-based approach introduces a latent Gaussian variable for each discrete one, and the conditional dependency structure of the full Gaussian vector is assumed to represent the corresponding structure of the observed variates~\citep{hofling2009estimation}.
The probit-based approach adopts a related approach in terms of inferring the conditional independence structure of mixed variables, but interprets discrete variables as thresholded transformations of underlying latent Gaussian variables~\citep{albert1993bayesian}.
We refer to the former model as copula-GGM and to the latter as probit-GGM.



Statistical inference about GGMs and MGMs has garnered significant attention across multiple disciplines.
For GGMs, a broad range of computational techniques has been developed for efficient structure learning,
aiming to address the challenges posed by high-dimensional datasets and complex dependency structures.
In a frequentist setting, a prominent approach for GGMs involves regularized maximum likelihood estimation or regression-based methods to infer the precision matrix, which subsequently determines the underlying graphical structure.
Notably, \citet{friedman2008sparse} introduced the graphical lasso, a computationally efficient algorithm with theoretical guarantees.
Numerous extensions and refinements of graphical lasso have been  explored~\citep{liu2009nonparanormal, xue2012regularized}.
The direction taken in this work is toward Bayesian posterior simulation, which necessitates the specification of priors for both the graph structure and the precision matrix.
When adopting the conjugate G-Wishart prior for the precision matrix given a  graph~\citep{roverato2002hyper},
posterior inference becomes challenging for conventional MCMC schemes  due to (among other factors) the intractability of a normalizing constant of the G-Wishart law for general non-decomposable graphs.
To address this issue, a series of advanced MCMC schemes have been proposed~\citep{wang2012efficient, cheng2012hierarchical, hinne2014efficient, lenkoski2013direct, mohammadi2015bayesian, mohammadi2023accelerating}.
For instance, \cite{hinne2014efficient} introduced the Double Conditional Bayes Factor (DCBF)  utilizing the exchange algorithm \citep{murray2012mcmc},
which circumvents the normalizing constant issue by leveraging auxiliary variable techniques to facilitate \emph{exact} posterior sampling (in the sense that no analytical approximation or Monte Carlo estimation of the involved normalizing constant(s) are utilized).
Building on this methodology, \cite{van2022g} developed the WWA algorithm, 
integrating delayed acceptance techniques~\citep{christen2005markov},
informed proposals~\citep{Dobra01122011} and local balancing for improved mixing and convergence~\citep{zanella2020informed}.
Specifically, rather than employing a uniform proposal distribution for the graph space,
the algorithm incorporates more informative proposals while utilizing locally balanced proposals designed for discrete spaces to enhance graph-exploration efficiency.
Additionally, a delayed acceptance step is introduced to accelerate computational efficiency by rapidly rejecting low-probability samples.
Beyond MCMC methods defined in discrete-time, continuous-time ones, such as Birth-Death MCMC (BDMCMC)~\citep{mohammadi2015bayesian, mohammadi2017bayesian}, have also emerged as competitive alternatives for posterior inference in GGMs. 
By leveraging a continuous-time birth-death process, wherein new states are always accepted with a waiting time determined by birth and death rates, 
BDMCMC efficiently navigates the graph space, demonstrating superior performance in Bayesian structure learning. 

Recent developments in Bayesian MGMs have further expanded the scope of GMs to accommodate heterogeneous data types, 
including continuous, ordinal, binary and count variables. 
In particular, conditionally specified modeling frameworks combined with spike-and-slab priors have been shown to offer flexibility in capturing complex dependence structures, 
including both positive and negative interactions, even in the presence of zero-inflation or missing data~\citep{GALIMBERTI2024105323, 10.1214/25-BA1557}. 
These advances highlight the growing need for scalable and efficient posterior simulation algorithms capable of exploring large and complex  graph spaces for mixed data. 
Related Bayesian work in genomic data integration and psychometrics has extended GGMs to ordinal data using  Bayes factors and model averaging for principled edge selection and structural uncertainty quantification~\citep{10.1214/22-AOAS1701, marsman2025bayesian}.

We summarize the main contribution of this work as follows: 
\begin{itemize}
\item[(a)]
We propose the Mixed Graph WWA (henceforth abbreviated to M-WWA) methodology for posterior simulation in MGMs
and conduct a comprehensive performance evaluation of our method.
Specifically, we develop two MCMC posterior simulation methods,
copula-WWA and probit-WWA, tailored for two key classes of MGMs, copula-MGMs and probit-MGMs,
respectively, building upon the WWA framework.
For copula-WWA, 
we employ a Gibbs sampler to generate latent variables from their full conditional, which corresponds to a truncated Gaussian distribution~\citep{d2007extending}.
The WWA algorithm is then used to sample the graph and the precision matrix given the  latent variates.
The probit-WWA algorithm first estimates the involved cut-off points using ordinal linear regression. Subsequently, we employ a Gibbs sampler to generate the latent variables~\citep{albert1993bayesian}, which are then used in the WWA simulation steps for posterior sampling. 
\item[(b)] We investigate in detail the performance of M-WWA algorithms against BDMCMC ones, initially on simulated datasets. 
\item[(c)] We apply M-WWA on the real PAM$50$ breast cancer gene expression dataset~\citep{parker2009supervised}, 
where the inferred MGMs reveal meaningful dependencies between gene expression profiles and cancer subtypes. Again, results from M-WWA are contrasted with those from BDMCMC.
\end{itemize}
\begin{rmk}
We note that the  approach of first estimating the cut-off points before proceeding to the identification of the graph has been used in a frequentist setting \citep{suggala2017ordinal} for  multivariate ordinal onservations (involving a sub-optimal use of bivariate likelihoods to estimate individual elements of the correlation matrix, before applying a sparsity operation) but has yet to be applied in a Bayesian one, to the best of our knowledge.
\end{rmk}
This article is structured as follows. Section~\ref{sec: mgm} introduces the notation and provides preliminary background on MGMs,
including detailed discussions on GGMs, copula-MGMs, and probit-MGMs.
Section~\ref{sec: mwwa} outlines the fundamentals of the WWA algorithm and presents the detailed MCMC methods for MGMs, including the probit-WWA and copula-WWA algorithms, under the general M-WWA framework.
Section~\ref{sec: numerical_exp} provides a thorough performance evaluation of our proposed methods using both synthetic and real-world data.
We conclude this paper and discuss future directions in Section~\ref{sec: diss}.

\section{Mixed Graphical Models}
\label{sec: mgm}
\subsection{Gaussian Graphical Models}
GGMs assume that observations follow a multivariate normal distribution 
that is Markovian w.r.t.~an underlying undirected graph.
Formally, let $G=(V, E)$ be an undirected graph, with
$V=\{1, \ldots, p\}$, $p\ge 1$, denoting the set of nodes, and $E \subseteq \{(i,j)\in V\times V: i<j\}$ 
representing the set of edges present in graph $G$.
We are given $n\ge 1$ iid observations,  $\mathcal{Y} =\{\mathbf{Y}^{(i)}\}_{i=1}^{n}$, $n\ge 1$,
each corresponding to a $p$-dimensional vector $\mathbf{Y}=(Y_1,\ldots, Y_p)^{\top}$, 
with  $\mathbf{Y}^{(i)}\sim \mathcal{N}_p(0, K^{-1})$, so that the mean is zero and $K$ is a symmetric positive-definite   precision matrix.
The pairwise Markov property states that if there is no edge between nodes $j, k$, i.e.~$(j, k)\notin E$,
then the corresponding random variables $Y_{j}$ and $Y_{k}$ are conditionally independent
given all remaining variables, i.e.,
$Y_{j}\perp Y_{k}|Y_{V\backslash\{j, k\}}$.
A standard result in the Gaussian setting shows that the pairwise Markov property holds
if and only if the corresponding entry in the precision matrix is zero, i.e.:
\begin{align*}
K_{j, k} = K_{k,j} = 0 \text{ for all } (j, k) \notin E,\quad  j < k.
\end{align*}
The above implies that the precision matrix $K$ of a GGM constrained by the graph $G$ belongs to the cone of symmetric positive definite matrices, denoted $M^{+}_{G}$, with entries of zeros at the positions corresponding to absence of edges. 

The likelihood of the observed data $\mathcal{Y}$ is given as follows: 
\begin{align*}
    p(\mathcal{Y}|K, G) = const\cdot |K|^{n/2}\exp\big\{-\tfrac{1}{2}\text{tr}(KS)\big\},
\end{align*}
where $S:=\sum_{i=1}^{n}\mathbf{Y}^{(i)}(\mathbf{Y}^{(i)})^{\top}$.
To construct Bayesian GGMs, we introduce prior distributions for both the graph $G$ and the precision matrix $K$ -- denoted $p(G)$ and $p(K|G)$ respectively. 
Several options are available for $p(G)$, with a simple one being the the uniform law on all graphs. See the next paragraph for more details.
A widely used prior for $p(K|G)$
is the G-Wishart distribution $\mathcal{W}_{G}(\delta, D)$,
which serves as a conjugate prior for GGMs~\citep{roverato2002hyper}.
The G-Wishart law is parameterized by the degrees of freedom $\delta > 2$, and a positive-definite rate matrix $D$,
with its pdf given by: 
\begin{align}
    f(K|G) = \frac{1}{I_{G}(\delta, D)}\,|K|^{\delta/2 - 1}\exp{\{-\tfrac{1}{2}\text{tr}(K^{\top}D)\}}, \qquad K\in M^{+}_{G},
\end{align}
for shape parameter $\delta>2$, and strictly positive-definite rate matrix $D$.  
Note that $I_{G}(\delta, D)$ is a normalizing constant ensuring a proper density.
Due to conjugacy, the posterior distribution $p(K|G,\mathcal{Y})$ is a G-Wishart one, with $K|G, \mathcal{Y}\sim \mathcal{W}_{G}(\delta^{*}, D^{*})$,
where the updated parameters are $\delta^{*} = \delta + n$ and $D^{*} = D + S$.
For decomposable graphs, the normalizing constant 
$I_{G}(\cdot,\cdot)$ can be computed explicitly~\citep{muirhead2009aspects}.
For general non-decomposable graphs, $I_{G}(\cdot, \cdot)$ lacks a closed-form expression, and this is a key point which adds computational challenges when carrying out posterior inference for $G,K$.
To address this complication, various approximations for $I_{G}(\cdot, \cdot)$ have been proposed, including Monte Carlo methods and Laplace approximations~\citep{atay2005monte, lenkoski2011computational}.

Extensive research has been conducted on prior distributions for graphs in the context of Bayesian GMs~\citep{carvalho2009objective, jones2005experiments, scott2006exploration, scutari2013prior}.
In this work, we focus on two widely used priors: the uniform prior and the sparse graph prior,
though our proposed Bayesian framework is applicable to arbitrary graph priors.
When prior knowledge about the graph structure is limited,
a low-information prior can be adopted which specifies that all graphs are equally probable.
This uniform prior is given by
$P(G) = \frac{1}{|\mathcal{G}|}$, where $|\mathcal{G}|$ is the cardinality of the space of possible graphs.
Sparse graph priors are often preferred to enforce parsimony,
reflecting the belief that real-world networks tend to have relatively few edges compared to the total number of possible connections. 
To maintain an expected number (a-priori) of edges of order $\mathcal{O}(p)$,
the edge inclusion probability can be set to be  $\mathcal{O}(\frac{1}{p})$.
This formulation ensures that, on average, the number of edges grows linearly with the number of nodes, thus maintaining sparsity in high-dimensional settings.

\begin{rmk}
To simplify the notation and presentation, we provide below the description of MGMs and of corresponding MCMC algorithms as if all $p$ variables were discrete. Certainly, 
in the typical setting of mixed models when continuous observations are also present, the corresponding variates among the total of $p$ ones will not require a special treatment.
\end{rmk}

\subsection{Probit- and Copula-Based Mixed Graphical Models}
Copula- and probit-based MGMs (also referred to as copula-MGMs and probit-MGMs in this paper) incorporate the structure of GGMs within hierarchical model structures to accommodate mixed data types.
These approaches allow discrete and continuous variables to be jointly modeled by introducing latent continuous variables governed by a GGM.
The observed variables are then derived from these latent representations through appropriate link functions or marginal transformations.

In probit-MGMs, each discrete observed variable is assumed to arise from an underlying continuous latent one~\citep{chib1998analysis}.
Specifically, the $j$-th component of the $i$-th observation of a discrete variable,  $Y^{(i)}_{j} \in \{1, \ldots, l_j\}$, $l_j \ge 1$,
is modeled using an ordered probit formulation,
where the corresponding latent variable $Z^{(i)}_j$ is standardized and mapped to ordered sub-intervals via an unknown threshold vector $\mathbf{C}_{j} = (C_{j, 1}, \ldots, C_{j, l_j-1})$ as follows:
\begin{align}
&\tilde{Z}^{(i)}_{j} = \frac{Z^{(i)}_j}{\sqrt{(K^{-1})_{j,j}}}, \nonumber \\
&Y^{(i)}_j = \sum_{k=1}^{l_j} k \cdot \mathbb{I}\,\big[\,C_{j, k-1} < \tilde{Z}^{(i)}_{j} < C_{j, k}\,\big],
\label{eq: probit}
\end{align}
with $C_{j,0} \equiv -\infty$, $C_{j,l_j} \equiv +\infty$.
Here, $K$ is the precision matrix of the multivariate normal law on the latent space,
and $\mathbb{I}\,[\,\cdot\,]$ the indicator function.
This framework allows for flexible modeling of  discrete outcomes, with a conditional independence structure encoded in the latent variables and captured by the precision matrix $K$.
See~\cite{chib1998analysis} for a foundational treatment of Bayesian probit modeling.

Copula-based MGMs rely on a latent Gaussian copula representation for mixed data. In particular, 
following the framework of~\cite{d2007extending}, an observed discrete variable $Y^{(i)}_j$ 
is again linked to an underlying latent Gaussian one,  through an appropriate marginal transformation.
The model separates the marginal distributions from the dependence structure by transforming the latent Gaussian variables using the corresponding univariate inverse cdf. That is, we set:
\begin{align}
\widetilde{Z}^{(i)}_{j} &= \frac{Z^{(i)}_j}{\sqrt{(K^{-1})_{j,j}}}, \nonumber \\
Y^{(i)}_j &= F_j^{-1} \left( \mathsf{\Phi}(\widetilde{Z}^{(i)}_j) \right), 
\label{eq: copula}
\end{align}
where $F_j(\cdot)$ is the marginal cdf of $Y_j$
and $F_j^{-1}(\cdot)$ its generalized inverse which is well-defined for discrete supports.
Function $\mathsf{\Phi}(\cdot)$ denotes the cdf of the  univariate standard normal distribution.
Through this transformation, the joint distribution of the observed variables is specified via a Gaussian copula,
while allowing for arbitrary marginal distributions~\citep{d2007extending}. Full details are provided in the next section.
The construct facilitates flexible modeling of mixed-type data within a coherent probabilistic framework while retaining the interpretability of the latent GGM structure.

\section{WWA Algorithm and its Extension to MGMs}
\label{sec: mwwa}
\subsection{Posterior Distribution for Mixed Gaussian Graphical Models}
\begin{figure}[!th]
\centering
\includegraphics[width=0.80\linewidth]{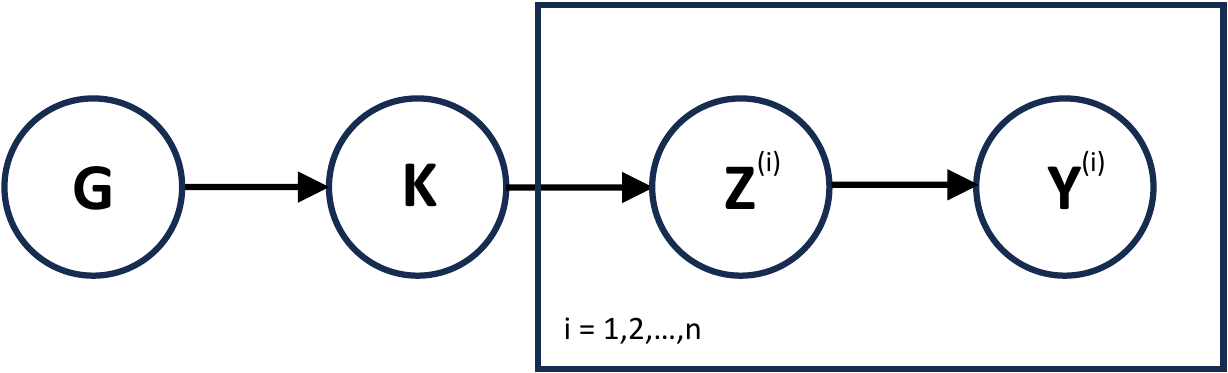}
\caption{Graphical representation of an MGM incorporating latent Gaussian variables.
A GGM generates latent variables $\mathbf{Z}^{(i)}$,
which are subsequently transformed into observed mixed variables $\mathbf{Y}^{(i)}$ via structured mappings such as probit- or copula-based transformations.
The rectangular plane indicates replication over $n$ independent samples.}
\vspace{0.3cm}
\label{fig:mgm_viz}
\end{figure}

We begin by introducing the notational conventions used in the sequel.
For each variable $j \in \{1, \dots, p\}$, we denote by $\phi_j(\cdot)$ the the deterministic transformation that maps the latent Gaussian variable $Z_j$ to the observed variable $Y_j$, i.e.:
\[\phi_j: Z_j \mapsto Y_j,\qquad j=1,\ldots, p.\]
The collection of all latent variables across the $n$ iid samples is denoted by $\mathcal{Z} = \{\mathbf{Z}^{(i)}\}_{i=1}^{n}$,
where each $\mathbf{Z}^{(i)}  = (Z_1,\ldots, Z_p)^{\top}\in \mathbb{R}^p$ represents the latent vector corresponding to the $i$-th observation.
Upon considering the conditional independence structure of an MGM (see Figure~\ref{fig:mgm_viz}),
we note that the precision matrix $K$ and the graph $G$ are independent of the observed data $\mathcal{Y} = \{\mathbf{Y}^{(i)}\}_{i=1}^{n}$, given the full set of latent variables $\mathcal{Z}$.
That is, the full conditional distribution of $(G,K)$ writes as follows:
\begin{align}
p(G, K \mid \mathcal{Z}, \mathcal{Y}) = p(G, K \mid \mathcal{Z})
\propto p(G)\,p(K \mid G)\,p(\mathcal{Z} \mid K),
\end{align}
where $p(G)$ is the prior on the graph,
$p(K | G)$ is the G-Wishart prior on $K$ given $G$,
and $p(\mathcal{Z} |K)$ is the Gaussian likelihood of the latent variables.
Bayesian structure learning of the graph $G$ and the associated precision matrix $K$ can now be achieved via MCMC methods.
A natural strategy is to employ a  Metropolis-within-Gibbs sampler that  alternates iteratively between the following two steps:
\begin{itemize}
    \item[(i)] Update of the latent variables $\mathcal{Z}$:\\
    One
    updates the latent variables $\mathcal{Z}$ conditioned on the observed data $\mathcal{Y}$ and the current state of $(K,G)$ and the particular choice of transformation functions $\{\phi_j(\cdot)\}_{j=1}^{p}$.
    Since the mapping $\phi_{j}(\cdot)$ from $Z_j$ to $Y_j$ is deterministic but non-invertible (in the  discrete setting of interest), this step typically involves truncated Gaussian sampling within 
    sub-intervals consistent with the observed data.
    \item[(ii)] Update of the graph $G$ and the precision matrix $K$:\\ Conditionally on the latent variables $\mathcal{Z}$ and observations $\mathcal{Y}$, one updates the graph $G$ and the corresponding precision matrix $K$ via an Metropolis-Hastings step that preserves the full conditional law of $(G,K)$.
    This step leverages the fact that such a full conditional is also treated in MCMC methods developed for standard GGMs.
\end{itemize}

This iterative scheme defines a blocked Metropolis-within-Gibbs sampler, where latent variables act as auxiliary components that bridge discrete observations and an underlying continuous Gaussian structure, enabling efficient and flexible posterior inference in mixed-type data settings.

\subsection{Graph and Precision Matrix Update via WWA}
To carry out an update that preserves the full conditional $p(G, K|\mathcal{Z})$,
we utilize the WWA algorithm of \cite{van2022g}.
WWA is an inference engine that will be at the core of both the copula-WWA and probit-WWA methods mentioned previously.
We begin by detailing the WWA procedure, which integrates recent advances in structure learning over GGMs via an efficient and \emph{exact} MCMC framework.
The WWA algorithm can be thought of as a major upgrade of the DCBF sampler \citep{hinne2014efficient}, 
and incorporates three key innovations to deliver fast convergence and computational efficiency:
\begin{itemize}
\item[(i)] Informed proposals leveraging approximated posterior ratios and local balancing.
    \item[(ii)] Exchange MCMC, based on a Cholesky transform, to overcome the intractability of the normalizing constant $I_{G}(\delta,D)$.
    \item[(iii)] Delayed Acceptance (DA) to avoid unnecessary computation in low-probability regions.
\end{itemize}
For completeness, we briefly present all above three main components of WWA. In what follows, let $G$ denote the current graph and $\widetilde{G}$ the proposed graph differing by a single edge (denoted $e$) from $G$. The discussion treats simultaneously both the addition and deletion of the edge $e$.
The nodes are re-labeled so that edge $e$ connects the nodes 
$\{p-1,p\}$. Let $\Phi^e$ denote the Cholesky factor of matrix $K^e$, 
i.e.~$K^e = (\Phi^e)^{\top}\Phi^e$ for the upper triangular matrix $\Phi^e$. 
The addition of superscript $e$ in the notation emphasizes that the relevant quantities are now considered after the above-mentioned re-labeling has taken place. 
We also define variate $\Phi^{e}_{-f}$ which serves to disentangle the intricate connection between the state spaces of $G$ and $K^e$ (equivalently, $\Phi^e$), that is: 
\begin{align*}
\Phi^{e}_{-f} = \Phi^e \backslash\{ \Phi^e_{p-1,p},\Phi^e_{p,p} \}.
\end{align*}
For given $G=(V,E)$, $|E|$ is the number of edges of $G$.
Let $\mathrm{nbd}(G)$ denote the neighborhood of graph $G$, i.e.~the set of graphs that differ from $G$ only by one edge. Due to conjugacy under the G-Wishart, we have  $K|G,\mathcal{Z}\sim \mathcal{W}_{G}(\delta^\ast, D^{\ast})$, where:
\begin{equation}
\label{eq:post_par}
\delta^{\ast} = \delta + n, \qquad D^{\ast} = D + \sum_{i=1}^{n}\mathbf{Z}^{(i)}(\mathbf{Z}^{(i)})^{\top}.
\end{equation}

\subsubsection*{Informed-Proposal Part}
On discrete spaces, \cite{zanella2020informed} sets up the locally-balanced framework for incorporating information from the likelihood in the Metropolis-Hastings proposal. In particular, for given target posterior $\pi(\cdot)$ and current discrete-valued position $x$, one has the proposal:
\begin{align}
Q(\widetilde{x}|x) = \tfrac{1}{c_g(x)} \cdot  g\Big\{ \frac{\pi(\widetilde{x})}{\pi(x)} \Big\}\,q(\widetilde{x}|x), \qquad \widetilde{x}\in \mathrm{nbd}(x) \label{eq:local},
\end{align}
for a baseline simple proposal $q(\cdot|\cdot)$, a normalizing constant $c_g(x)>0$,  and a user-specified definition of the neighborhood $\mathrm{nbd}(x)$ typically based on the geometry of the given discrete space. 
The balancing function $g(\cdot)$ should satisfy $g(t)=t\cdot g(1/t)$, so that small steps in an increasingly large space will be accepted w.p.~1 in the limit (in an analogue of the Random-Walk Metropolis behavior on continuous spaces). Standard choices are $g(t)=\sqrt{t}$, $g(t)=1/(1+t)$; we use the latter form in the applications. Importantly, computations over all $\widetilde{x}\in \mathrm{nbd}(x)$ are parallelizable.

In our setting, we are required to treat the jointly discrete and continuous space of $(G,K)$. Conditionally on $K$, WWA utilizes of the following informed proposal: 
\begin{align}
&Q(\widetilde{G}|G,K,\mathcal{Z}) = c(G,K)\cdot\, 
g\bigg\{ \frac{p_u(\widetilde{G}, \Phi^{e}_{-f}|\mathcal{Z})}{p_u(G, \Phi^{e}_{-f}|\mathcal{Z})}\times \widehat{I_G/I_{\widetilde{G}}}\bigg\}\cdot q(\widetilde{G}|G),\qquad  \widetilde{G}\in\mathrm{nbd}(G),
\label{eq:WWA_prop}
\end{align}
for $\widehat{I_G/I_{\widetilde{G}}}$ being a well-studied approximation of the ratio of the true normalizing constant; recall the $e$ is the single edge that separates $G=(V,E)$ and $\widetilde{G}=(V,\widetilde{E})$. In particular, following \cite{mohammadi2023accelerating} for the standard choice $D=I_p$, we set:  
\begin{align}
\label{eq:estimate_nc}
\widehat{I_G/I_{\widetilde{G}}} 
= \bigg\{  \frac{\Gamma(\frac{\delta+d_{\widetilde{G}}}{2})}{2\sqrt{\pi}\,\Gamma(\frac{\delta+d_{\widetilde{G}}+1}{2})} 
\bigg\}^{|\widetilde{E}|-|E|},
\end{align}
where $d_{\widetilde{G}}$ is the number of paths of length two linking the
endpoints of edge $e$, and $\Gamma(\cdot)$ is the Gamma function. The ratio of unnormalized densities appearing in (\ref{eq:WWA_prop}) is analytically available, that is \citep{cheng2012hierarchical}: 
\begin{equation}
\label{eq:tractable}
\frac{p_u(\widetilde{G}, \Phi^{e}_{-f}|\mathcal{Z})}{p_u(G, \Phi^{e}_{-f}|\mathcal{Z})} = \frac{p(\widetilde{G})}{p(G)}\cdot \mathcal{G}(\Phi_{-f}^e, D^{\ast,e})^{|\widetilde{E}|-|E|}.
\end{equation}
Here, we have defined, for general $p\times p$ matrix $S$:
\begin{align*}
&\mathcal{G}(\Phi_{-f}^e, S) = \Phi^e_{p-1,p-1}\cdot \sqrt{\tfrac{2\pi}{S_{p,p}}}\cdot \exp\Big( \tfrac{1}{2}S_{p,p}(\phi_0-\mu)^2 \Big), 
\\ & \mu = \Phi^e_{p-1,p-1} S_{p-1,p}/S_{p,p}, \qquad \phi_0 = - \frac{1}{\Phi^e_{p-1,p-1}}\sum_{l=1}^{p-2}\Phi_{l,p-1}\Phi_{l,p}.
\end{align*}
\subsubsection*{Exchange-Algorithm Part}

The discussion below requires samples from the G-Wishart distribution, $\mathcal{W}(\delta,D)$. For this purpose
we employ the direct sampler of~\cite{lenkoski2013direct}. 

\begin{figure}[!th]
\vspace{0.5cm}
\centering
\includegraphics[width=0.50\linewidth]{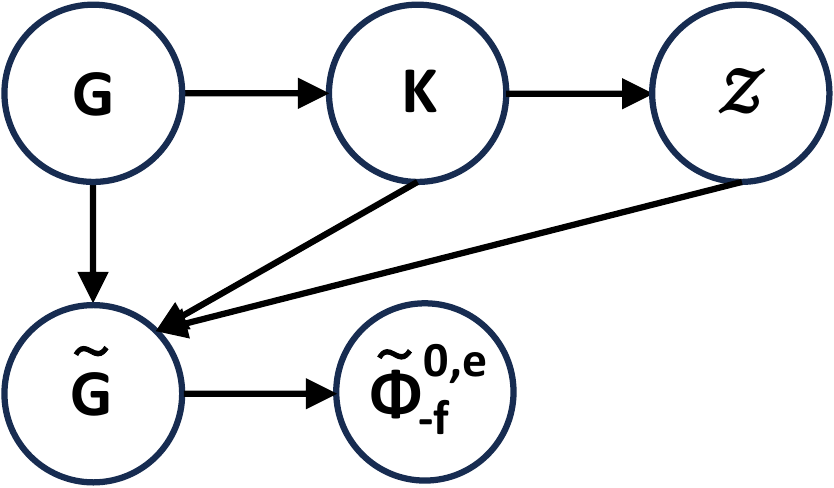}
\caption{The DAG of the model one obtains after augmenting the GGM corresponding to the joint law of $(G,K,\mathcal{Z})$ with the proposed $\widetilde{G}$ from (\ref{eq:WWA_prop}) and the auxiliary variable $\widetilde{\Phi}^{0,e}_{-f}$. \\[-0.1cm]
An exchange algorithm step will generate a proposal for the discrete-continuous pair $(G,K)$ by suggesting the deterministic switch $G\longleftrightarrow\widetilde{G}$ alongside a random update for $K$.}
\label{fig:WWA_DAG}
\end{figure}

We adopt the underlying idea of the exchange algorithm \citep{murray2012mcmc} and introduce an auxiliary precision matrix $\widetilde{K}^{0}$ from the prior of the GGM model  $p(\widetilde{K}^{0}|\widetilde{G})$. The construct gives rise also
to the elements $\widetilde{K}^{0,e}$, $\widetilde{\Phi}^{0,e}$, $\widetilde{\Phi}^{0,e}_{-f}$.
Thus, together with the proposal set up in (\ref{eq:WWA_prop}), we have now formulated a model with auxiliary components having the Directed Acyclic Graph (DAG) shown in Figure~\ref{fig:WWA_DAG}.

Summarizing the joint law of the augmented model in Figure~~\ref{fig:WWA_DAG} writes as follows: 
\begin{align}
&p(G,K,\widetilde{G},\widetilde{\Phi}_{-f}^{0,e}|\mathcal{Z}) = p(G,K|\mathcal{Z})\cdot Q(\widetilde{G}\,|G,K,\mathcal{Z})\cdot p(\widetilde{\Phi}_{-f}^{0,e}|\widetilde{G}). \label{eq:ext_target}
\end{align}
The exchange idea on the joint space of $(G,K)$ now applies a deterministic switch for $G$ (as with the standard exchange algorithm) alongside a random update for $K$ to enforce alignment between proposed graph and precision matrix.
In particular, the MH-proposal with target the distribution in (\ref{eq:ext_target}) is as follows:
\begin{itemize}
\item[(a)] $(G,\widetilde{G})$-components: Exchange (deterministically) $G\longleftrightarrow\widetilde{G}$.
\item[(b)] $K$-component: Propose updating $K$ by $\widetilde{K}^e(\equiv (\widetilde{\Phi}^e)^{\top}\widetilde{\Phi}^e)$, constructed as follows (let $q(\widetilde{K}|G,K,\widetilde{G})$ denote the proposal density): 
\begin{itemize}
\item[(i)] Set $\widetilde{\Phi}_{-f}^{e} = \Phi_{-f}^{e}$ and $\widetilde{\Phi}_{p,p}^{e} = \Phi_{p,p}^{e}$;
\item[(ii)] Sample   $\widetilde{\Phi}^{e}_{p-1,p}\sim p(\widetilde{\Phi}^{e}_{p-1,p} | \widetilde{\Phi}_{-f}^{e},\widetilde{G},\mathcal{Z})$.
\end{itemize}
\end{itemize}
In terms of the law appearing in step (b)(ii) above, \cite{van2022g} show that: 
\begin{equation}
\label{eq:sample_p}
\begin{aligned}
&\widetilde{\Phi}^{e}_{p-1,p} | \widetilde{\Phi}_{-f}^{e},\widetilde{G},\mathcal{Z} \sim 
\mathcal{N}_1\Big(  \frac{-\widetilde{\Phi}^e_{p-1,p-1}D^{\ast,e}_{p-1,p}}{D_{p,p}^{\ast,e}},\frac{1}{D_{p,p}^{\ast,e}}  \Big), \qquad  e\in \widetilde{E}, \\
&\widetilde{\Phi}^{e}_{p-1,p} | \widetilde{\Phi}_{-f}^{e},\widetilde{G},\mathcal{Z}  = -\frac{1}{\widetilde{\Phi}_{p-1,p-1}}\sum_{l=1}^{p-2}\widetilde{\Phi}_{l,p-1}^e\widetilde{\Phi}_{l,p}^e,\qquad e\notin \widetilde{E}.
\end{aligned}
\end{equation}
Another conditional law involving an element of $\Phi^e$ as the above and which is used in the final algorithm is the following: 
\begin{align}
D_{p,p}^{\ast, e}(\Phi_{p,p}^{e})^2\,|\,G, \Phi_{p-1,p}^e, \Phi_{-f}^e,\mathcal{Z} \sim  \chi^2(\delta^{\ast}).
\label{eq:sample_pp}
\end{align}
See \cite{van2022g} for the derivation.

A corresponding acceptance probability (denoted $1\wedge R^{\mathrm{informed}}_{\mathrm{exchange}}$) delivers  detailed balance w.r.t.~$p(G,K,\widetilde{G},\widetilde{\Phi}_{-f}^{0,e}|\mathcal{Z})$.
We have that: 
\begin{align}
R^{\mathrm{informed}}_{\mathrm{exchange}} = \frac{p(\widetilde{G},\widetilde{K}^{e},G,\widetilde{\Phi}_{-f}^{0,e}|\mathcal{Z})\,q(K|\widetilde{G},\widetilde{K},G)}{p(G,K,\widetilde{G},\widetilde{\Phi}_{-f}^{0,e}|\mathcal{Z})\,q(\widetilde{K}|G,K,\widetilde{G})}.
\label{eq:exchange_ratio}
\end{align}
The machinery of the exchange algorithm leads to a cancellation of the normalizing constants in (\ref{eq:exchange_ratio}) following the presence of the auxiliary component $\widetilde{\Phi}_{-f}^{0,e}$. Indeed, the nominator in (\ref{eq:exchange_ratio}) involves the term $\{I_{\widetilde{G}}(\delta,D)\cdot I_{G}(\delta,D)\}^{-1}$, and the same also applies for the denominator. 
The analytically tractable expression for (\ref{eq:exchange_ratio}) is derived in detail in \cite{van2022g} and is given as follows: 
\begin{align}
\label{eq:R_formula}
R^{\mathrm{informed}}_{\mathrm{exchange}} = 
\frac{p(\widetilde{G})}{p({G})}\cdot \Big\{   
\frac{\mathcal{G}(\Phi^e_{-f},D^{\ast,e})}
{\mathcal{G}(\widetilde{\Phi}^{0,e}_{-f},D^{e})}
\Big\}^{|\widetilde{E}|-|E|}\cdot \frac{Q(G|\widetilde{G},\widetilde{K},\mathcal{Z})}{Q(\widetilde{G}|G,K,\mathcal{Z})}.
\end{align}


\subsubsection*{Delayed-Acceptance Part}

We have so far set up proposal $(G, K)\longrightarrow (\widetilde{G},\widetilde{K})$ and an acceptance probability giving rise to a Markov transition that preserves $p(G,K|Y)$.
%
%
WWA also adds a Delayed Acceptance (DA) step, aimed at reducing the number of instances when a sample from $\mathcal{W}_G$ is requested. DA adds an extra test to $(\widetilde{G},\widetilde{K})$ before the pair is as proposal.  The acceptance probability that WWA utilizes for promoting a proposal $(\widetilde{G},\widetilde{K})$ is equal to $1\wedge R_{\mathrm{DA}}$ where:
\begin{align}
\label{eq:RDA}
R_{\mathrm{DA}} = \frac{p_u(\widetilde{G},\Phi_{-f}^e|\mathcal{Z})\cdot Q(G\,|\widetilde{G},\widetilde{K}^{e},\mathcal{Z})}{p_u(G,\Phi_{-f}^e|\mathcal{Z})\cdot Q(\widetilde{G}\,|\,G,K,\mathcal{Z})} \times \widehat{ I_G/I_{\widetilde{G}}}.
\end{align}
Note that $R_{_{\mathrm{DA}}}$ is calculated via expressions (\ref{eq:WWA_prop}), (\ref{eq:estimate_nc}) and (\ref{eq:tractable}). 
The underlying idea is that $R_{\mathrm{DA}}$ resembles  the MH-ratio of a standard exchange algorithm that switches $G\longleftrightarrow \widetilde{G}$, with the precision matrix that needs to be treated in a careful manner in the exchange-algorithm part now being simply fixed to current and proposed values $K$, $\widetilde{K}$ that align with the corresponding current and proposed graphs $G$, $\widetilde{G}$ respectively. 

By avoiding the (expensive) sampling the G-Wishart law required for the generation of $\widetilde{\Phi}^{0,e}_{-f}$ and the evaluation of the full posterior for proposals likely to be rejected,
DA allows the WWA algorithm to focus computational effort on high-potential moves, thus 
leading to substantial gains in efficiency, 
particularly in large to high-dimensional graphical models.

\subsubsection*{Complete WWA Algorithm}
Following the standard DA machinery \citep{christen2005markov}, if $(\widetilde{G}, \widetilde{K})$ is promoted, the final acceptance probability writes as $1\wedge R$, where:   
\begin{align}
\label{eq:R}
R = \frac{1\wedge R_{\mathrm{DA}}^{-1}}{1\wedge R_{\mathrm{DA}}}\times R_{\mathrm{exchange}}^{{\mathrm{informed}}}.
\end{align}
Recall that $R_{\mathrm{exchange}}^{{\mathrm{informed}}}$ is analytically available as in (\ref{eq:R_formula}).
 From standard properties of DA, the two-step acceptance procedure provides detailed balance 
w.r.t.~$p(G,K|\mathcal{Z})$. Combining the three components described above,
we summarize the full WWA methodology in Algorithm~\ref{alg:wwa}.

\begin{algorithm}[H]
\label{alg:wwa}
\caption{WWA Algorithm -- A Single MCMC Iteration}
\KwIn{Current latent variables $\mathcal{Z}$; current graph $G = (V,E)$; balancing function $g(t)=1/(1+t)$; no of attempted edge updates, $n_e\ge 1$.}\vspace{0.3cm}
   \begin{itemize}
   \item[1.] Sample the precision matrix  $K|G, \mathcal{Z}\sim \mathcal{W}_{G}(\delta^{*}, D^{*})$
   for $(\delta^{*},D^{*})$ as in (\ref{eq:post_par}). \\[0.3cm]
   \item[2.] \For{$n_{\text{e}}$ iterations}{ \vspace{0.25cm}
       (a) Sample graph $\widetilde{G}$ from the informed proposal $Q(\widetilde{G}\,|\,G, K,\mathcal{Z})$ specified  in~(\ref{eq:WWA_prop}). \\[0.2cm]
       (b) Reorder nodes, i.e.~rearrange $K\leftrightarrow K^e$, $D\leftrightarrow D^e$, $D^\star\leftrightarrow D^{\star,e}$. \\[0.25cm]
       (c) Update $\Phi_{p,p}^{e}$ by sampling from ${\Phi}^{e}_{p,p}\,\big|\,G,{\Phi}^{e}_{p-1,p},{\Phi}^{e}_{-f},\mathcal{Z}$, using (\ref{eq:sample_pp}). \\[0.2cm]
       (d) Generate proposal $\widetilde{K}^e$ by sampling $\widetilde{\Phi}^{e}_{p-1,p}\sim p(\widetilde{\Phi}^{e}_{p-1,p}\,\big|\,\widetilde{G},\widetilde{\Phi}^{e}_{-f},\mathcal{Z})$ from (\ref{eq:sample_p}). \\[0.2cm]
       (e) Compute $Q(G\,|\widetilde{G},\widetilde{K},\mathcal{Z})$.\\[0.3cm]
       (f)  Promote $(\widetilde{G}, \widetilde{K})$ w.p. $1\wedge R_{\mathrm{DA}}$ for $R_{\mathrm{DA}}$ in (\ref{eq:RDA}). If $(\widetilde{G}, \widetilde{K})$ is promoted: \\[0.25cm]
       \begin{itemize}
       \item[-] Sample $\widetilde{K}^{0,e}\sim \mathcal{W}_{\widetilde{G}}(\delta, D^e)$ and  retrieve $\widetilde{\Phi}_{-f}^{0,e}$. \\[0.1cm]
       \item[-] Accept $(G,K)\leftarrow(\widetilde{G},\widetilde{K})$ w.p. $1\wedge R$, for $R$ in (\ref{eq:R}). \\[0.2cm]
       \end{itemize}
   }
 \end{itemize}
 \vspace{0.2cm}
 \KwOut{Next G.}
\end{algorithm}
\vspace{10px}

\subsection{Models \& Latent Variable Sampling}
In this section, we describe the latent sampling schemes for both the probit-WWA and copula-WWA algorithms. For the latent sampling of the copula transformation,
we utilize the idea proposed by~\cite{d2007extending} and described in detail in \cite{dobra2011copula}.

\subsubsection*{Copula-MGM}
The copula-MGM builds up via consideration of latent Gaussian variables as follows:
\begin{align}
    &\mathcal{Z}=(\mathbf{Z}^{(1)}, \ldots, \mathbf{Z}^{(n)}) \mid K \overset{\text{iid}}{\sim} \mathcal{N}_p(0, K^{-1}), \label{eq:copula-sampling} \\[0.3cm]
    &Y^{(i)}_{j} = F_j^{-1} \Big( \mathsf{\Phi}\big(\tfrac{Z^{(i)}_j}{\sqrt{(K^{-1})_{j,j}}}\big) \Big), \quad j=1,\ldots, p, \quad i=1,\ldots,n \nonumber.
\end{align}
Here, $F_j^{-1}(\cdot)$ denotes the (pseudo-)inverse of the unknown cdf of variable $Y_{j}$.
The approach sidesteps modeling of the individual $F_j$'s, utilizing instead only their non-decreasing properties.  
That is, we have the following relation:
\begin{align*}
Y^{(i)}_{j} < Y^{(k)}_{j}\;\Longrightarrow\; Z^{(i)}_ {j} < Z^{(k)}_{j}.
\end{align*}
%
Collectively, the above rule gives rise to a constraint set for the latent variables:
\begin{equation}
\label{eq:truncate}
\begin{aligned}
A(\mathcal{Y}):= \Big\{Z_j^{(i)}\in\mathbb{R}:
&\,\,L_j^{(i)}
< Z^{(i)}_{j}< U_j^{(i)},\,\, j=1,\ldots, p,\,\, i=1,\ldots,n\Big\},
\end{aligned}
\end{equation}
where we have defined: 
\begin{equation}
\label{eq: bd_est_copula}
L_{j}^{(i)} = \max_{k} \{ Z^{(k)}_{j} : Y^{(k)}_{j} < Y^{(i)}_{j} \}, \qquad 
U_{j}^{(i)} = \min_k \{ Z^{(k)}_{j} : Y^{(i)}_{j} < Y^{(k)}_{j} \}.
\end{equation}

The copula-MGM approach is based on the consideration of the following full conditional distribution:
\begin{equation}
p_{\mathrm{cop}}(\mathcal{Z}\,|\,K,G,\mathcal{Y})\,\propto\, p(\mathcal{Z}\,|\,K)\cdot 
\mathrm{P}\,[\,\mathcal{Z}\in A(\mathcal{Y})\,]. 
\label{eq:post_cop}
\end{equation}
Thus, to update $\mathcal{Z}$ based on the full conditional (\ref{eq:post_cop}) one applies a component-wise Gibbs sampler, with each of the $n\times p$ steps simulating from the truncated Gaussian laws:  
\begin{equation}
Z^{(i)}_{j} \mid K, \mathcal{Z}_{-j}^{(-i)}, \mathcal{Y} \,\sim\, 
\mathcal{N}_1\Big(- \sum_{l \neq j} \tfrac{K_{j,l}}{K_{j,j}}Z^{(i)}_{l}, \; \tfrac{1}{K_{j,j}}
\Big)\cdot \mathbb{I}\,[\,L_j^{(i)}
< Z^{(i)}_{j}<
U_j^{(i)}\,].
\end{equation}

\subsubsection*{Probit-MGM}

Probit-MGMs adopt the framework of cumulative link models \citep{agre:2013} or, more specifically, that of ordered probit models \citep{gree:10} to transform continuous variates into discrete ones.

We assume a threshold vector for the $j$-th variable,
$\mathbf{C}_{j}=(C_{j, 0}, \ldots, C_{j, l_j})$, $l_{j}\ge 1$, so that $-\infty=C_{j, 0} < C_{j, 1}\cdots < C_{j, l_j}=\infty$.
Here, $l_j$ denotes the cardinality of the discrete space.
Then, one sets an hierarchical cumulative link model by first defining the latent Gaussian variates: 
\begin{align}
\label{eq: probit_model}
&\mathcal{Z}=(\mathbf{Z}^{(1)}, \ldots, \mathbf{Z}^{(n)}) \mid K \overset{\mathrm{iid}}{\sim} \mathcal{N}_p(0, K^{-1}), \\
&\widetilde{Z}^{(i)}_{j} = \frac{Z^{(i)}_{j}}{\sqrt{(K^{-1})_{j,j}}}.
\end{align}
These are then transformed to the observed variables via: 
\begin{align*}
Y^{(i)}_{j} = \sum_{k=1}^{l_j} k \cdot \mathbb{I}\,[\,C_{j, k-1} < \widetilde{Z}^{(i)}_{j} < C_{j,k}\,].
\end{align*}
Full Bayesian treatment of the latent Gaussian variates and the thresholds $\mathbf{C}_{j}$, $1\le j\le p$, would give rise to an overly complex model. Instead, we fix the $\mathbf{C}_{j}$'s to their MLE estimates $\widehat{\mathbf{C}}_{j}$ and treat them as known, that is one considers separate log-likelihoods over $j$: 
\begin{align*}
&\ell(c_{j, 1:l_{j}-1}|\mathcal{Y}) = \sum_{i=1}^{n}\log \pi_{j}^{(i)},  \\
&\pi_{j}^{(i)} = \sum_{k=1}^{l_j}\mathbb{I}\,[\,Y^{(i)}_{j}=k\,]\cdot\big(\mathsf{\Phi}(c_{j, k}) - \mathsf{\Phi}(c_{j, k-1})
\big),
\end{align*}
and obtains the values (for ordered $c_{j,1}<\cdots <c_{j, l_j-1}$): 
\begin{align}
\widehat{C}_{j,1:l_{j}-1} = \arg \max_{c_{j,1:l_j-1}}\ell(c_{j, 1:l_j-1}|\mathcal{Y}).
\label{eq:mle_c}
\end{align}
Upon fixing $C_{j,1:l_j-1}$ at their estimates $\widehat{C}_{j,  1:l_{j}-1}$, and setting the boundary cut-points $C_{j,0}$ and $C_{j,l_j}$ to $-\infty$ and $+\infty$, respectively, 
one can derive the full conditional distributions of the latent variables $\mathcal{Z}=(\mathbf{Z}^{(1)},\ldots,\mathbf{Z}^{(n)})$. 
These full conditional distributions are truncated normal distributions, with the truncation bounds determined by the observed variables.

\subsubsection*{Full Conditionals for the two Classes of MGMs}

We summarize the algorithms that sample from the full conditionals of the latent variables $\mathbf{Z}^{(1)}, \ldots, \mathbf{Z}^{(n)}$ both for the probit- and the copula-MGM.


\begin{algorithm}[H]
\label{alg: latents_sampling_copula}
\caption{Gibbs Update of $\mathcal{Z}=(\mathbf{Z}^{(1)}, \ldots, \mathbf{Z}^{(n)})$ for  Copula-MGM.}
\KwIn{Data $\mathcal{Y}$, current precision matrix $K$ and current $\mathcal{Z}=(\mathbf{Z}^{(1)}, \ldots, \mathbf{Z}^{(n)})$.}

\vspace{0.1cm}
\hrule
\vspace{0.2cm}

\textbf{
Copula-MGM} \\[0.2cm]

\For{$j = 1,\ldots p$}{ \vspace{0.2cm}
    \For{$i = 1,\ldots,n$}{ \vspace{0.1cm}
        1.\,\, Determine the lower and upper truncation bounds, $L_{j}^{(i)}, U_j^{(i)}$, according to \eqref{eq: bd_est_copula}.\\[0.1cm]
        2. \,\,Update the latent variable $Z^{(i)}_{j}$ by sampling from the truncated Gaussian law:
        \begin{equation*}
        Z^{(i)}_{j} \mid K,\,\mathcal{Z}_{-j}^{(-i)},\,\mathcal{Y} \,\sim\, 
\mathcal{N}_1\Big(- \sum_{l \neq j} \tfrac{K_{j,l}}{K_{j,j}}Z^{(i)}_{l}, \; \tfrac{1}{K_{j,j}}
\Big)\cdot \mathbb{I}\,[\,L_j^{(i)}
< Z^{(i)}_{j}<
U_j^{(i)}\,].
        \end{equation*}
    }
}
\vspace{0.2cm}
\KwOut{Collection of realisations of $\mathcal{Z}$ to be used in subsequent Gibbs steps.}
\end{algorithm}

\begin{algorithm}[H]
\label{alg: latents_sampling_probit}
\caption{Gibbs Update of $\mathbf{Z}^{(1)}, \ldots, \mathbf{Z}^{(n)}$ for  Probit-MGM.}
\KwIn{Data $\mathcal{Y}$, thresholds $\mathbf{C}_{j}$, $j=1,\ldots,p$, current  $K$ and current $\mathcal{Z}$.}

\vspace{0.1cm}
\hrule
\vspace{0.2cm}
\textbf{
Probit-MGM} \\[0.2cm]
\For{$j=1,\ldots,p$}{\vspace{0.2cm}
    \For{$i = 1,\ldots,n$} {\vspace{0.2cm}
        1.\,\,Determine the threshold pair $C^{(i)}_{j, \mathrm{left}}$, $C^{(i)}_{j, \mathrm{right}}$  corresponding to  $Y^{(i)}_{j}$. Specifically,
        we set $C^{(i)}_{j, \text{left}} = C_{j, Y_{j}^{(i)}-1}$ and $C^{(i)}_{j, \text{right}} = C_{j, Y_{j}^{(i)}}$.
        \\[0.1cm] 
        2.\,\,Sample the latent variable $Z^{(i)}_{j}$ from the truncated Gaussian law: 
        \begin{align*}
        Z^{(i)}_{j}\mid K,\,Z^{(i)}_{-j},\,\mathcal{Y} \sim \, & \mathcal{N}\Big(- \sum_{l \neq j} \tfrac{K_{j,l}}{K_{j,j}}Z^{(i)}_{l}, \tfrac{1}{K_{j,j}}\Big)\cdot
        \\
        & \mathbb{I}\,\Big[\,\sqrt{(K^{-1})_{j,j}}\cdot C^{(i)}_{j, \mathrm{left}} < Z^{(i)}_{j} < \sqrt{(K^{-1})_{j,j}}\cdot C^{(i)}_{j, \mathrm{right}}\,\Big].
        \end{align*}
    }
}
\vspace{0.2cm}
\KwOut{Collection of realisations of $\mathcal{Z}$ to be used in subsequent Gibbs steps.}
\end{algorithm}



\subsection{M-WWA for Mixed Graphical Models}
The developed M-WWA algorithm alternates between latent-variable augmentation and graph-space exploration.
Given the current graph $G$ and precision matrix $K$, the latent variables $\mathcal{Z}$ are update via use of their full conditional distributions.
Given $\mathcal{Z}$, the WWA algorithm is applied to update $K,G$ via consideration of their own full conditional law.
Repeated application of these steps yields a Markov chain with the desired posterior as its stationary distribution.
The full M-WWA algorithm is summarized in Algorithm \ref{alg:M-WWA}. We note that that the constructed targets, say $p_{\mathrm{cop}}(G,K|\mathcal{Y})$ and $p_{\mathrm{probit}}(G,K|\mathcal{Y})$ will typically differ  from the true ideal posterior, $p(G,K|\mathcal{Y})$.

\begin{algorithm}[H]
\label{alg:M-WWA}
\caption{M-WWA Algorithm}
\KwIn{Data $\mathcal{Y}$; initial choices of graph $G = (V,E)$, matrix $K$ and latent variables $\mathcal{Z}$; number of MCMC iterations, $M\ge 1$.} \vspace{0.2cm}
 \For{$S\in 1,\ldots M$ }{\vspace{0.2cm}
   1. Update the latent variables $\mathcal{Z}$ making use of the appropriate truncated Gaussian laws, for the probit-based and copula-based approaches. \\[0.1cm]
   2. Run WWA conditional on the current $\mathcal{Z}$ to update $K$ and $G$.
 }
 \vspace{0.2cm}
\KwOut{Collection of $(M+1)$ MCMC samples targeting either $p_{\mathrm{cop}}(G,K|\mathcal{Y})$ or $p_{\mathrm{probit}}(G,K|\mathcal{Y})$.}
\end{algorithm}

\begin{rmk}
We note that once the likelihood-`type' component in the specification of the target posterior has been specified, both in the probit- and the copula-based approaches, then 
M-WWA is `exact' in the sense that it is based on Markov transitions that provably preserve the constructed posterior. This is mainly due to the fact that no approximation is used when treating the intractable normalising constant. For the algorithms that use the BDgraph methodology, such a property is not guaranteed. The works in \cite{mohammadi2015bayesian,mohammadi2023accelerating} do discuss incorporating the exchange idea within the BDgraph methodology, but no formal proof for the invariant distribution under such a scheme is provided, even if combining continuous-time dynamics (used by BDgraph) with exchange-algorithm steps (which are intrinsically of discrete-time nature) does require careful examination for the purposes of specification of an invariant distribution.
\end{rmk}

\vspace{10px}



\section{Numerical Experiments}
\label{sec: numerical_exp}
\subsection{Simulation Data Analysis}
We evaluate the performance of the four methods in terms of recovery of the true graph using the standard $F_{1}$-score~\citep{10.1093/bioinformatics/16.5.412}, defined as:
\begin{align*}
    F_{1} = \frac{2TP}{2TP + FP + FN},
\end{align*}
where $TP$, $FP$ and $FN$ denote the numbers of true positives, false positives, and false negatives, respectively.
In the simulation study, we compare the structure recovery performance of  four methods, 
namely copula-WWA, probit-WWA, copula-BD, and probit-BD, under a range of graph structures, dimensions, and sample sizes using the $F_{1}$-score.  Here copula-BD, and probit-BD denote the MCMC algorithms which use the BDgraph algorithm for carrying out the update from $G,K|\mathcal{Z}$ under the copula-MGM and probit-MGM models respectively.
For the BD-based methods, 
the intractable ratio of normalising constants required for the graph-transition rates 
is approximated using an analytical formula.

\begin{table}[!htbp]
\centering
\resizebox{\textwidth}{!}{%
\begin{tabular}{llcccc}
\hline
 & & Copula-WWA & Probit-WWA & Copula-BDMCMC & Probit-BDMCMC \\
\hline
\multicolumn{6}{l}{$p=10, n=50$} \\
cycle  & & 0.73 (0.02) & 0.57 (0.01) & 0.71 (0.00) & 0.53 (0.02) \\
star    & & 0.15 (0.00) & 0.00 (0.00) & 0.15 (0.00) & 0.00 (0.00) \\
random  & & 0.52 (0.03) & 0.38 (0.02) & 0.51 (0.03) & 0.43 (0.02) \\
cluster & & 0.44 (0.03) & 0.33 (0.01) & 0.47 (0.02) & 0.49 (0.02) \\
scale-free & & 0.50 (0.01) & 0.39 (0.02) & 0.48 (0.00) & 0.33 (0.00) \\
\hline
\multicolumn{6}{l}{$p=10, n=100$} \\
cycle  & & 0.86 (0.01) & 0.71 (0.02) & 0.85 (0.03) & 0.62 (0.06) \\
star    & & 0.27 (0.00) & 0.15 (0.00) & 0.27 (0.00) & 0.22 (0.09) \\
random  & & 0.57 (0.02) & 0.47 (0.01) & 0.63 (0.02) & 0.49 (0.03) \\
cluster & & 0.69 (0.02) & 0.53 (0.01) & 0.68 (0.02) & 0.53 (0.00) \\
scale-free & & 0.67 (0.00) & 0.65 (0.03) & 0.65 (0.03) & 0.63 (0.00) \\
\hline
\multicolumn{6}{l}{$p=20, n=100$} \\
cycle  & & 0.68 (0.04) & 0.45 (0.05) & 0.61 (0.00) & 0.41 (0.01) \\
star    & & 0.63 (0.02) & 0.58 (0.03) & 0.62 (0.02) & 0.56 (0.05) \\
random  & & 0.53 (0.02) & 0.48 (0.02) & 0.59 (0.02) & 0.49 (0.02) \\
cluster & & 0.65 (0.01) & 0.64 (0.02) & 0.65 (0.01) & 0.61 (0.04) \\
scale-free & & 0.54 (0.01) & 0.43 (0.02) & 0.50 (0.02) & 0.43 (0.06) \\
\hline
\multicolumn{6}{l}{$p=20, n=200$} \\
cycle  & & 0.62 (0.00) & 0.56 (0.05) & 0.63 (0.00) & 0.43 (0.04) \\
star    & & 0.71 (0.02) & 0.32 (0.07) & 0.66 (0.01) & 0.29 (0.05) \\
random  & & 0.63 (0.02) & 0.52 (0.02) & 0.66 (0.02) & 0.53 (0.02) \\
cluster & & 0.86 (0.01) & 0.66 (0.03) & 0.85 (0.01) & 0.69 (0.01) \\
scale-free & & 0.66 (0.01) & 0.73 (0.02) & 0.67 (0.02) & 0.74 (0.01) \\
\hline
\multicolumn{6}{l}{$p=50, n=250$} \\
cycle  & & 0.46 (0.03) & 0.24 (0.02) & 0.38 (0.01) & 0.21 (0.02) \\
star    & & 0.30 (0.05) & 0.11 (0.06) & 0.25 (0.03) & 0.11 (0.06) \\
random  & & 0.57 (0.01) & 0.48 (0.02) & 0.56 (0.00) & 0.47 (0.01) \\
cluster & & 0.64 (0.02) & 0.54 (0.01) & 0.61 (0.02) & 0.52 (0.01) \\
scale-free & & 0.56 (0.01) & 0.53 (0.01) & 0.50 (0.02) & 0.55 (0.04) \\
\hline
\multicolumn{6}{l}{$p=50, n=500$} \\
cycle  & & 0.64 (0.03) & 0.29 (0.02) & 0.58 (0.00) & 0.26 (0.00) \\
star    & & 0.47 (0.02) & 0.59 (0.05) & 0.45 (0.01) & 0.57 (0.10) \\
random  & & 0.65 (0.01) & 0.51 (0.00) & 0.63 (0.00) & 0.50 (0.02) \\
cluster & & 0.76 (0.01) & 0.53 (0.01) & 0.74 (0.01) & 0.53 (0.01) \\
scale-free & & 0.59 (0.01) & 0.73 (0.02) & 0.60 (0.01) & 0.77 (0.00) \\
\end{tabular}%
}
\caption{Comparison of 4 different MCMC methods by $F_{1}$-score under various choices for $p$, $n$. The table shows the average $F_1$-scores and (in parenthesis) the standard errors, across $30$ independent replications.}
\label{tab: results}
\end{table}

As shown in Table~\ref{tab: results}, the M-WWA methods exhibit competitive performance across a broad range of experimental settings.
In particular, copula-WWA consistently attains high 
$F_{1}$-scores and generally outperforms the corresponding copula-BD approach, 
demonstrating the advantage of the M-WWA methods for efficient exploration of the posterior graph space, especially in setting of relatively higher dimension.
The comparison between latent-variable representations shows that copula-based methods generally achieve better recovery accuracy than their probit-based counterparts, 
particularly for cycle, random, and cluster graph structures. 
Such an understanding indicates that flexible dependence modeling of the copula representation seem to be able to better capture complex interactions among mixed variables. 
Meanwhile, probit-based methods remain competitive for certain scale-free structures, suggesting that the optimal latent representation may depend on the underlying graph topology.
As the sample size increases, all methods exhibit improved $F_{1}$-scores, confirming a desirable statistical behavior with increasing data availability. 
Overall, the results demonstrate that the proposed M-WWA framework provides an efficient and reliable approach for Bayesian structure learning in MGMs.

\begin{figure}[!htbp]
\begin{subfigure}{.49\textwidth}
\centering
\includegraphics[width=1\linewidth]{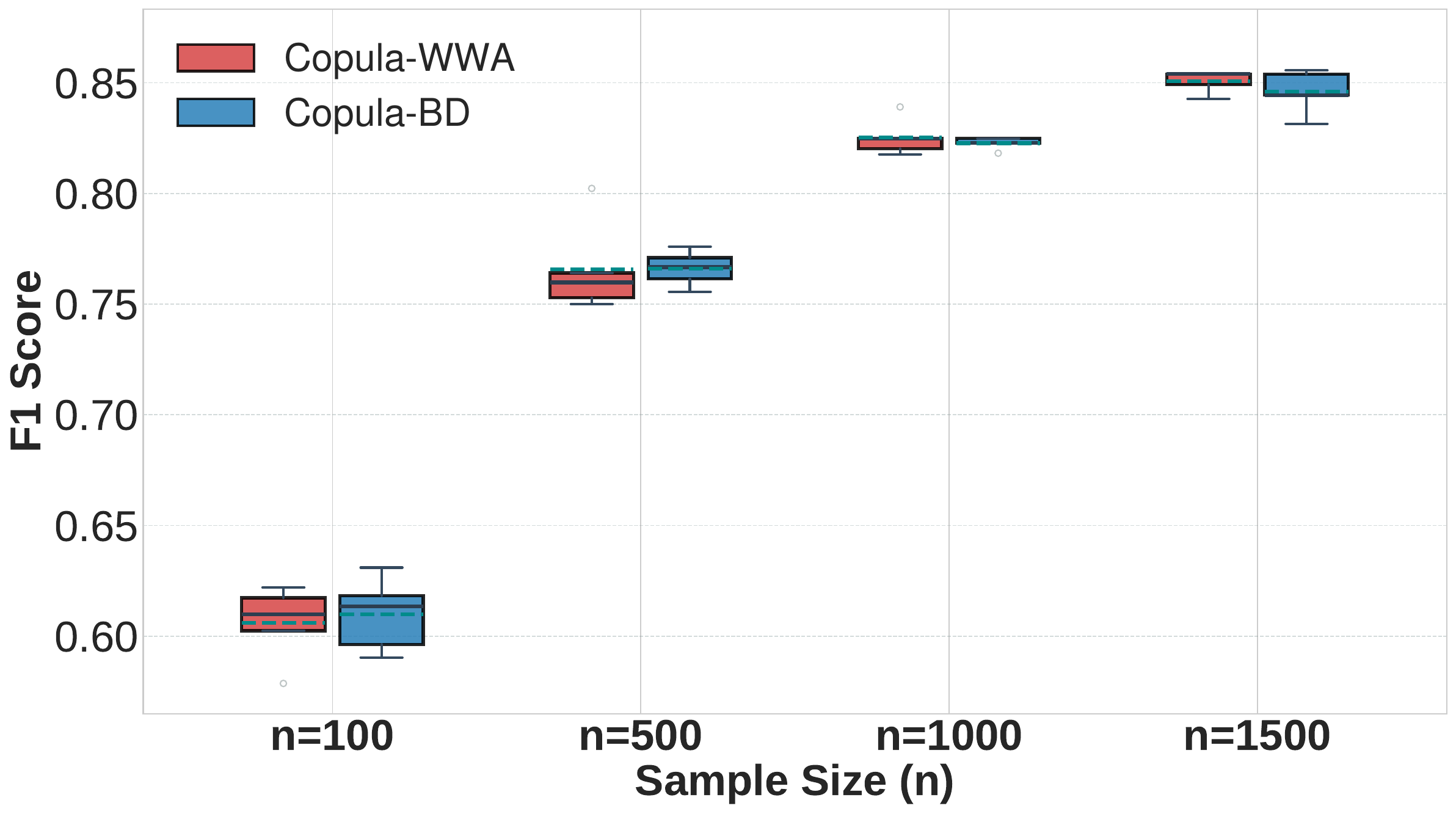}
\caption{Random Graph with $p = 20$ nodes.}
\label{fig:f1_score_p_20_random}
\end{subfigure}
\begin{subfigure}{.49\textwidth}
\centering
\includegraphics[width=1\linewidth]{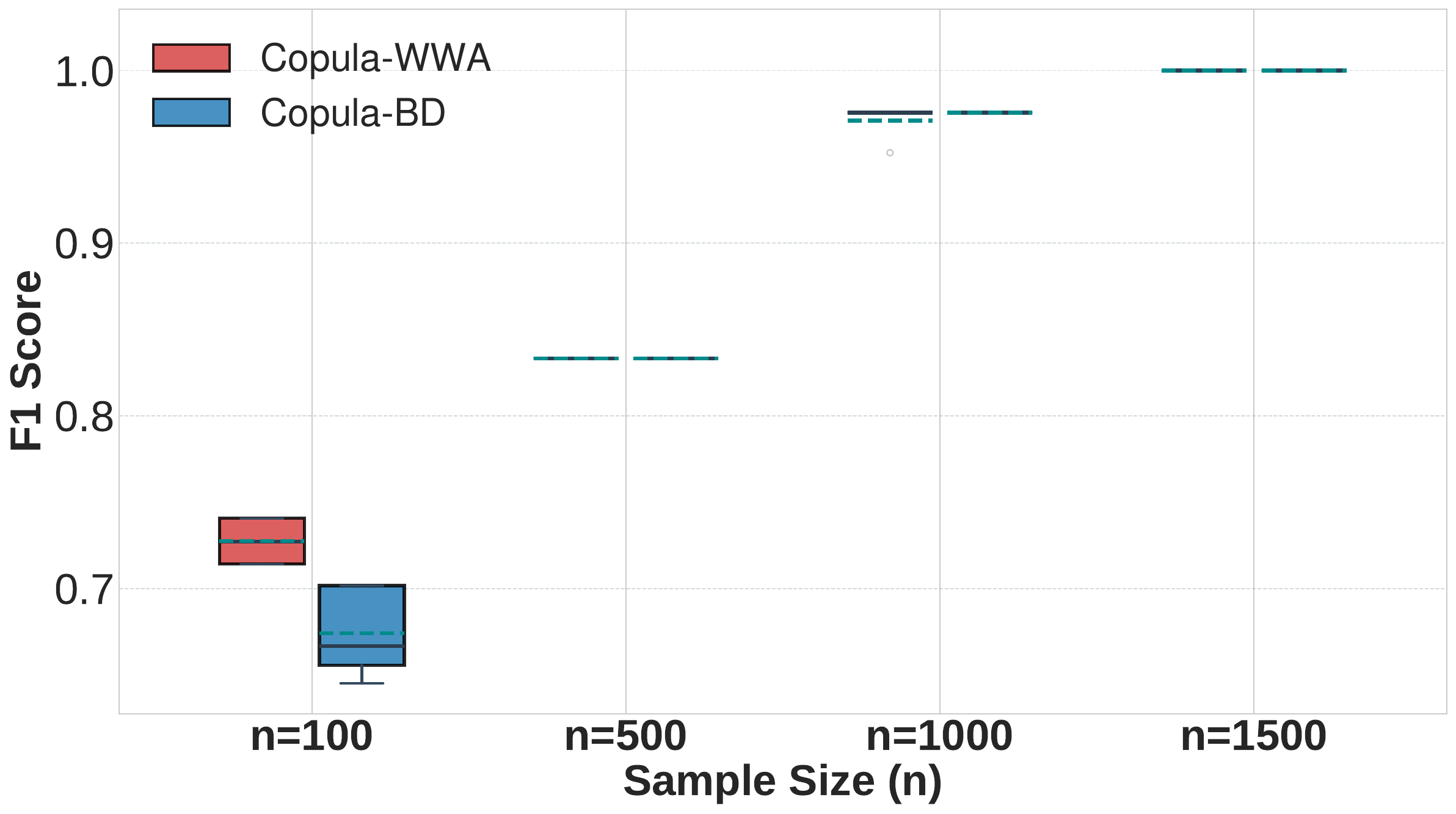}
\caption{Cycle Graph with $p = 20$ nodes.}
\label{fig:f1_score_p_20_circle}
\end{subfigure}
\caption{
Boxplots of the $F_{1}$-scores for the copula-WWA and copula-BD algorithms under the random and cycle graph structures with varying sample sizes.
}
\label{fig: cbd}
\end{figure}

Figures~\ref{fig:f1_score_p_20_random}-\ref{fig:f1_score_p_20_circle} further illustrate the $F_{1}$-scores across different sample sizes for MGMs
with a random graph as underlying structure with $p=20$ nodes and a cycle graph as underlying structure with $p=20$ nodes, respectively.
For random graph architectures with $p=20$ dimensions, copula-WWA maintains competitive $F_{1}$ performance relative to copula-BD across the full spectrum of sample sizes, 
exhibiting comparable mean accuracy while demonstrating robustness in edge recovery even under availability of limited samples.
The copula-WWA algorithm substantially outperforms copula-BD in small-sample regimes for cycle graphs with $p=20$ nodes, achieving a mean $F_1$-score of $0.727$ compared to copula-BD's $0.674$ at $n=100$, 
while both methods attain perfect accuracy at $n=1,500$, indicating copula-WWA's superior efficiency in structured graph recovery under data scarcity.
Detailed simulation settings are provided in Appendix~\ref{app: simu_details}.

\subsection{Real Data Analysis}
\label{sec: real}
To address the molecular heterogeneity of breast cancer, \cite{parker2009supervised} introduced the PAM$50$ gene dataset, corresponding to a standardized panel of $50$ genes that enables reliable classification of breast tumors into five major molecular subtypes: `luminal A', `luminal B', `basal-like', `HER2-enriched', and `normal-like'.
The dataset consists of $772$ breast cancer patients with known subtype labels: $416$ luminal A, $141$ luminal B, $135$ basal, $46$ HER2-enriched, and $34$ normal-like.
This classification, derived from gene expression profiling, has become a widely adopted framework in both clinical and research settings, offering a robust tool for studying subtype-specific differences in prognosis, treatment response, and survival outcomes.

\begin{figure}[!htbp]
\begin{subfigure}{.48\textwidth}
\centering
\includegraphics[width=0.9\linewidth]{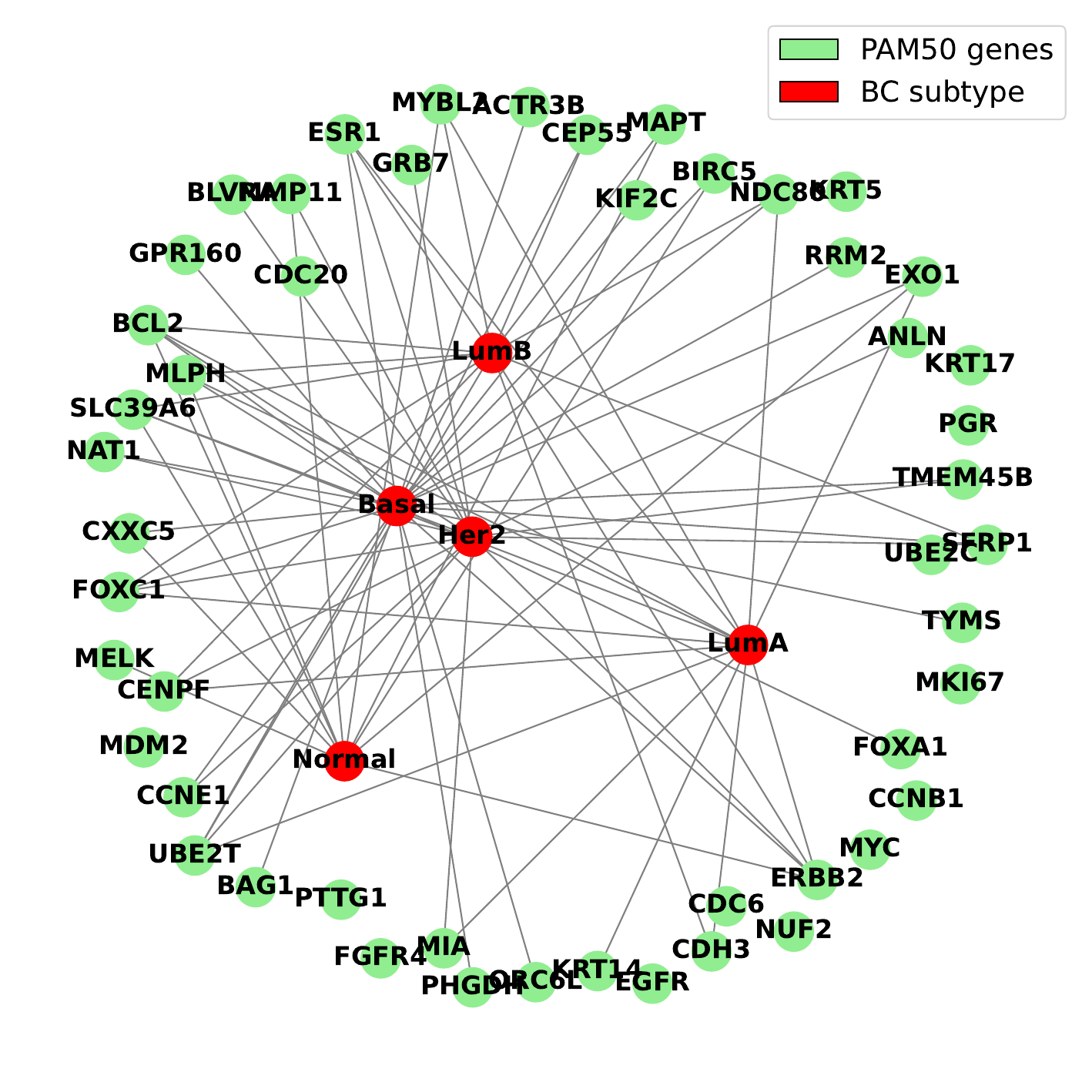}
\caption{Copula-WWA}
\label{fig: gsub_cwwa_pam50}
\end{subfigure}
\begin{subfigure}{.48\textwidth}
\centering
\includegraphics[width=0.9\linewidth]{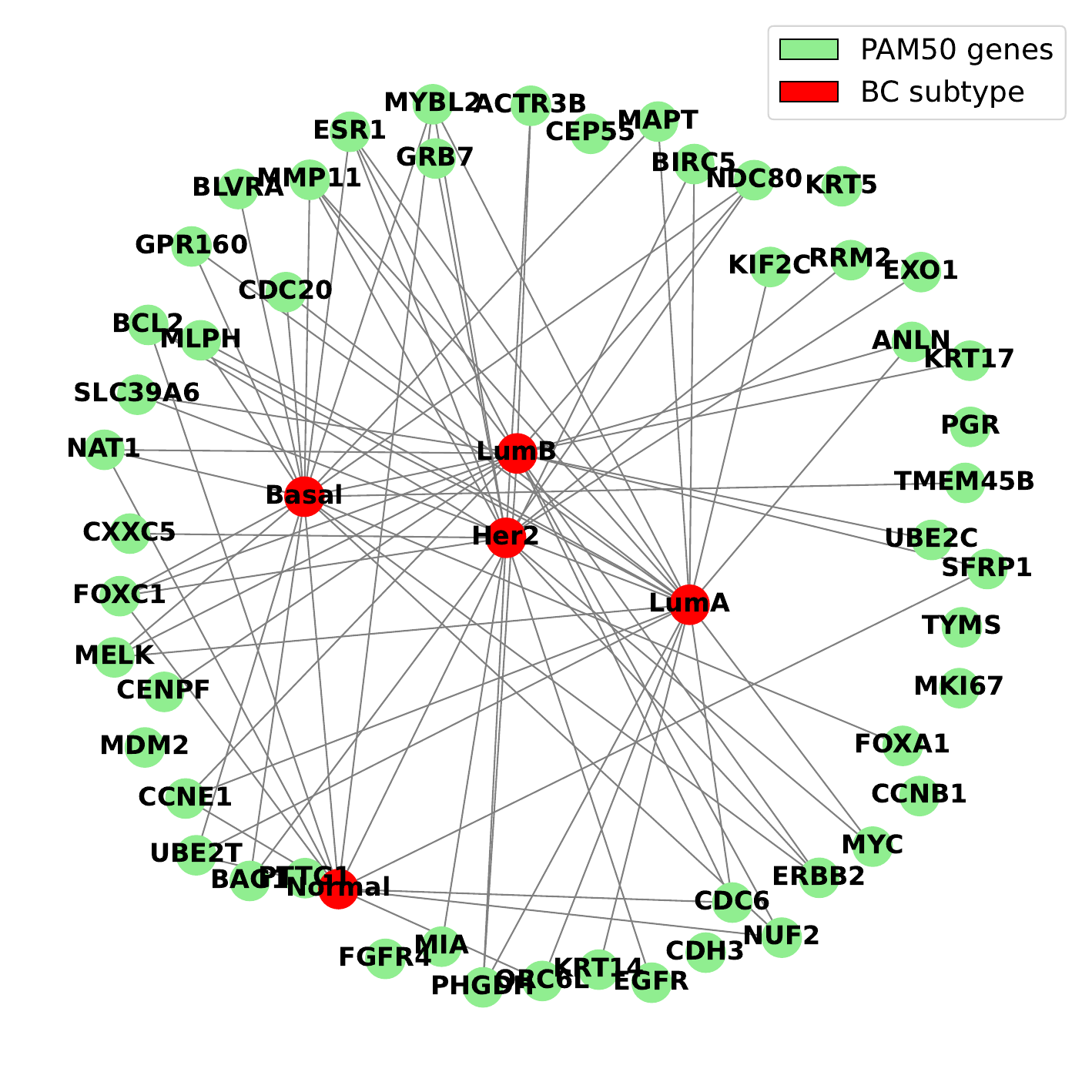}
\caption{Copula-BD}
\label{fig: gsub_cbd_pam50}
\end{subfigure}\\
\begin{subfigure}{.48\textwidth}
\centering
\includegraphics[width=0.9\linewidth]{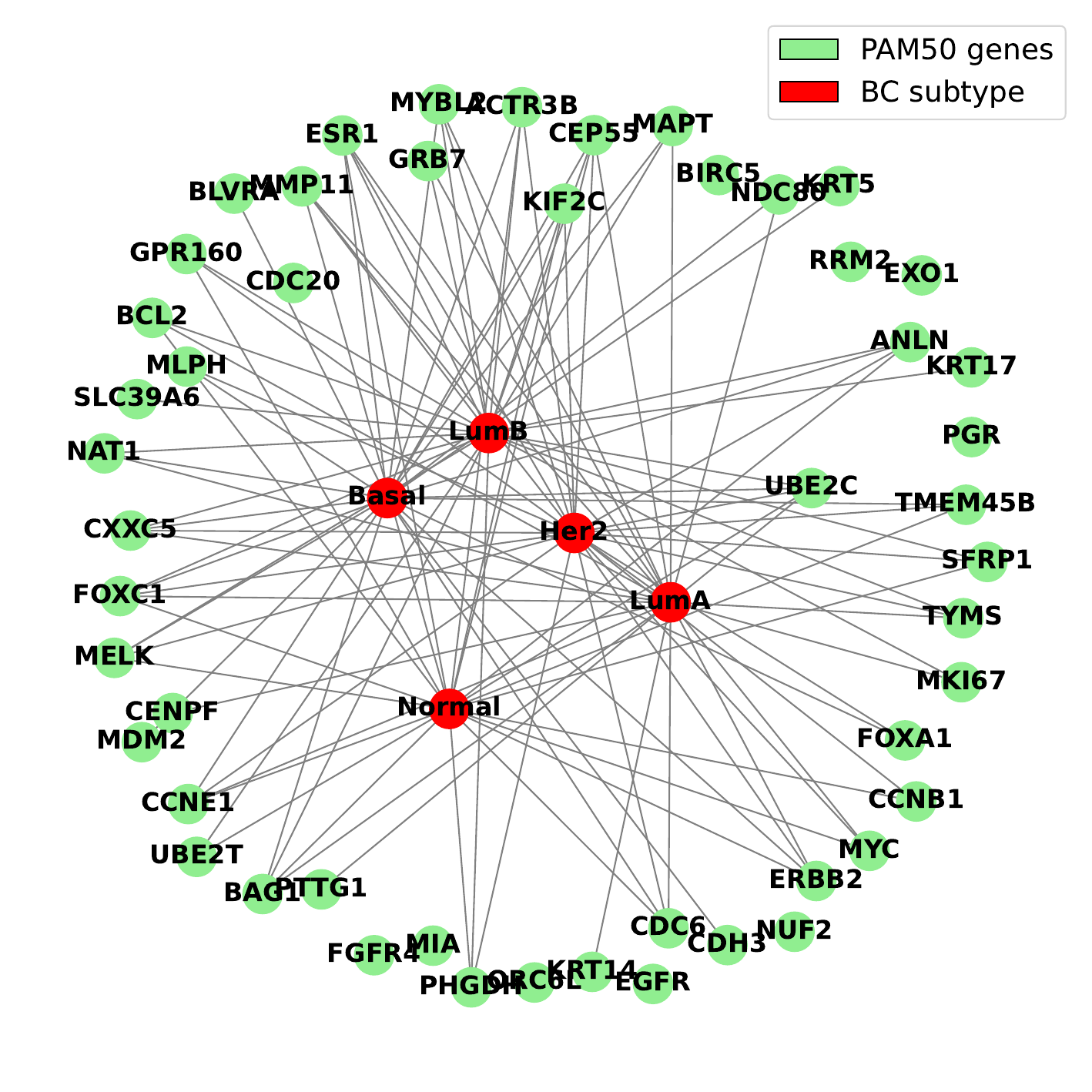}
\caption{Probit-WWA}
\label{fig: gsub_pwwa_pam50}
\end{subfigure}
\begin{subfigure}{.48\textwidth}
\centering
\includegraphics[width=0.9\linewidth]{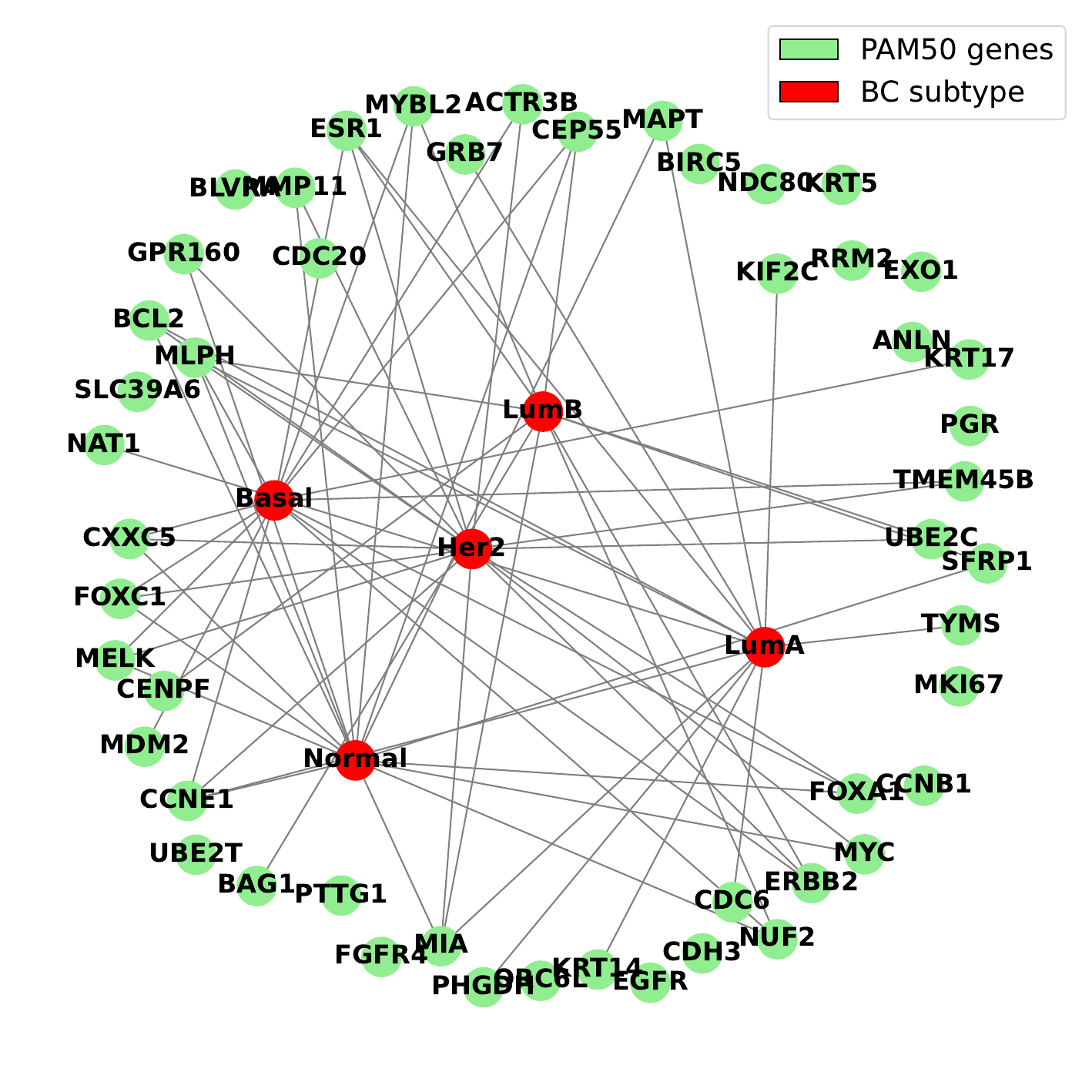}
\caption{Probit-BD}
\label{fig: gsub_pbd_pam50}
\end{subfigure}
\caption{Estimated MGMs for the PAM$50$ breast cancer dataset. 
The upper-left panel corresponds to copula-WWA, 
the upper-right panel to copula-BD, 
the lower-left panel to probit-WWA, and the lower-right panel to probit-BD. 
}
\label{fig: gsub_pam50}
\end{figure}

For the analysis of the PAM$50$ dataset, we applied the copula-WWA, probit-WWA, copula-BD, and probit-BD algorithms to estimate MGMs that capture dependencies between gene expression levels and breast cancer subtypes.
Specifically, we log-transformed the expression levels of the $50$ PAM$50$ genes to approximate Gaussian distributions,
and incorporated them alongside five 0/1 binary variables correposponding to absence/presence for each of the five breast cancer subtypes.
The estimated MGMs obtained by the four methods are shown in Figure~\ref{fig: gsub_pam50}. 

\begin{figure}[!htbp]
\begin{subfigure}{.48\textwidth}
\centering
\includegraphics[width=0.9\linewidth]{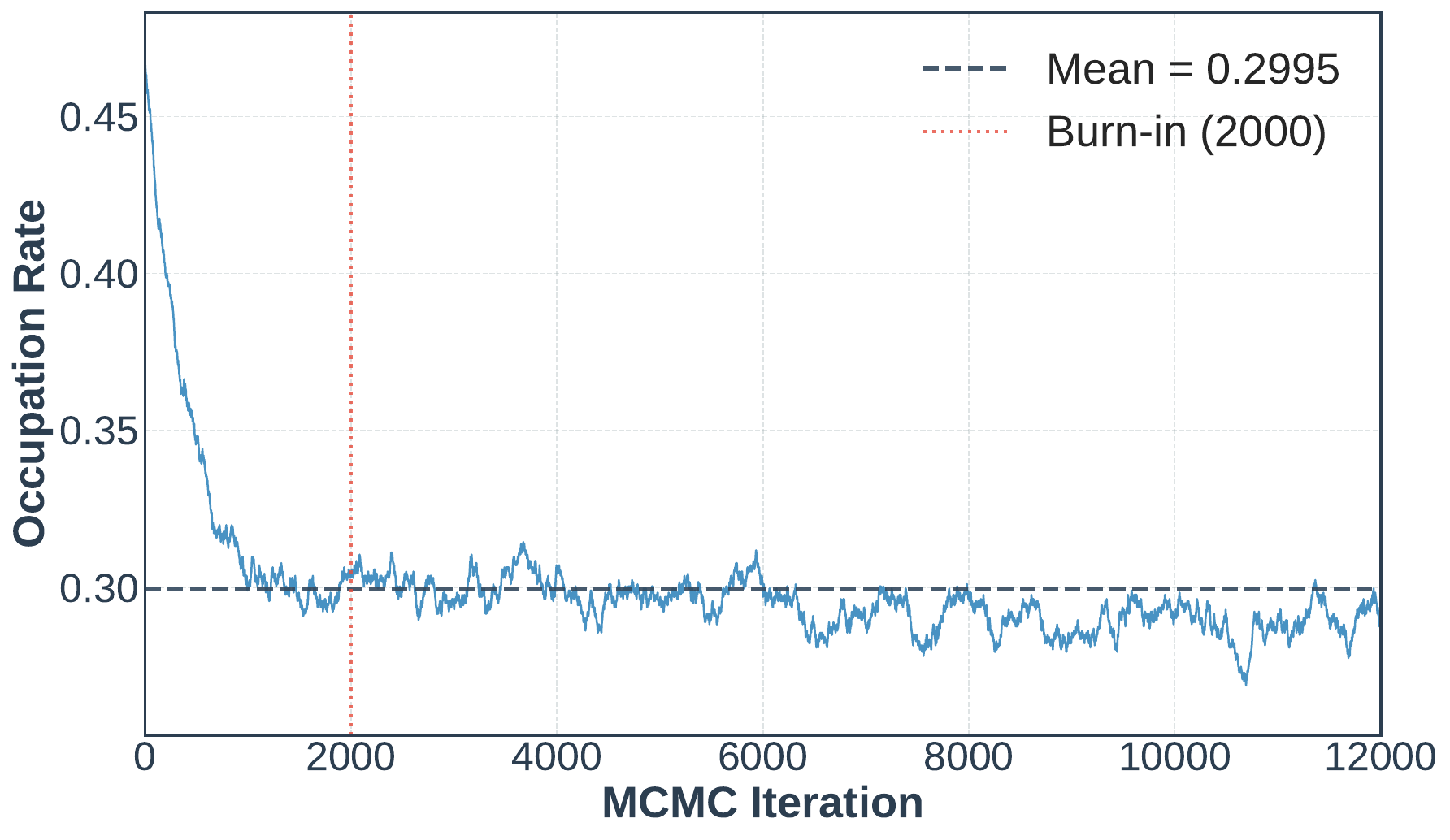}
\label{fig: occ_cwwa_pam50}
\end{subfigure}
\begin{subfigure}{.48\textwidth}
\centering
\includegraphics[width=0.9\linewidth]{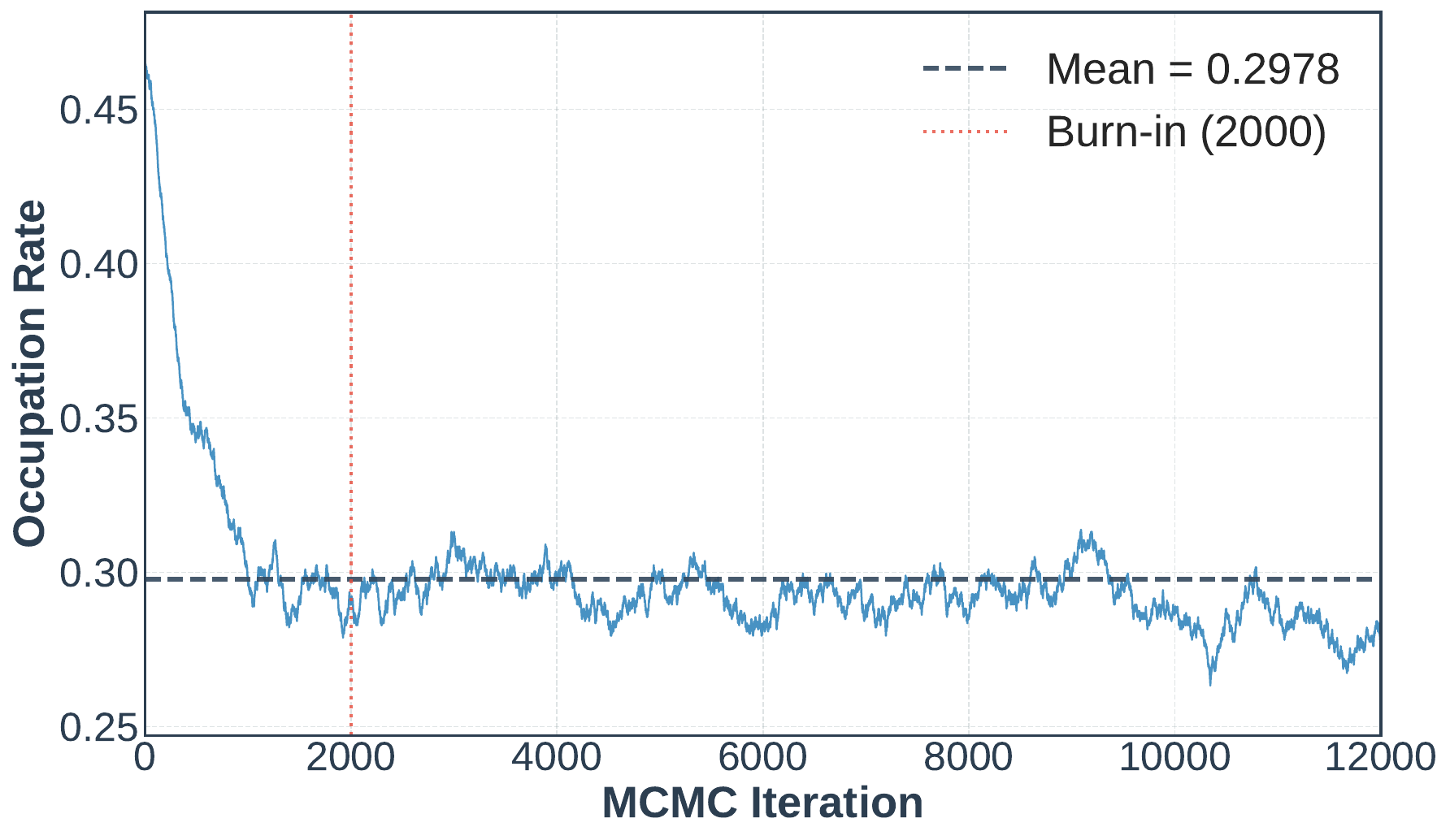}
\label{fig: occ_cbd_pam50}
\end{subfigure}\\
\begin{subfigure}{.48\textwidth}
\centering
\includegraphics[width=0.9\linewidth]{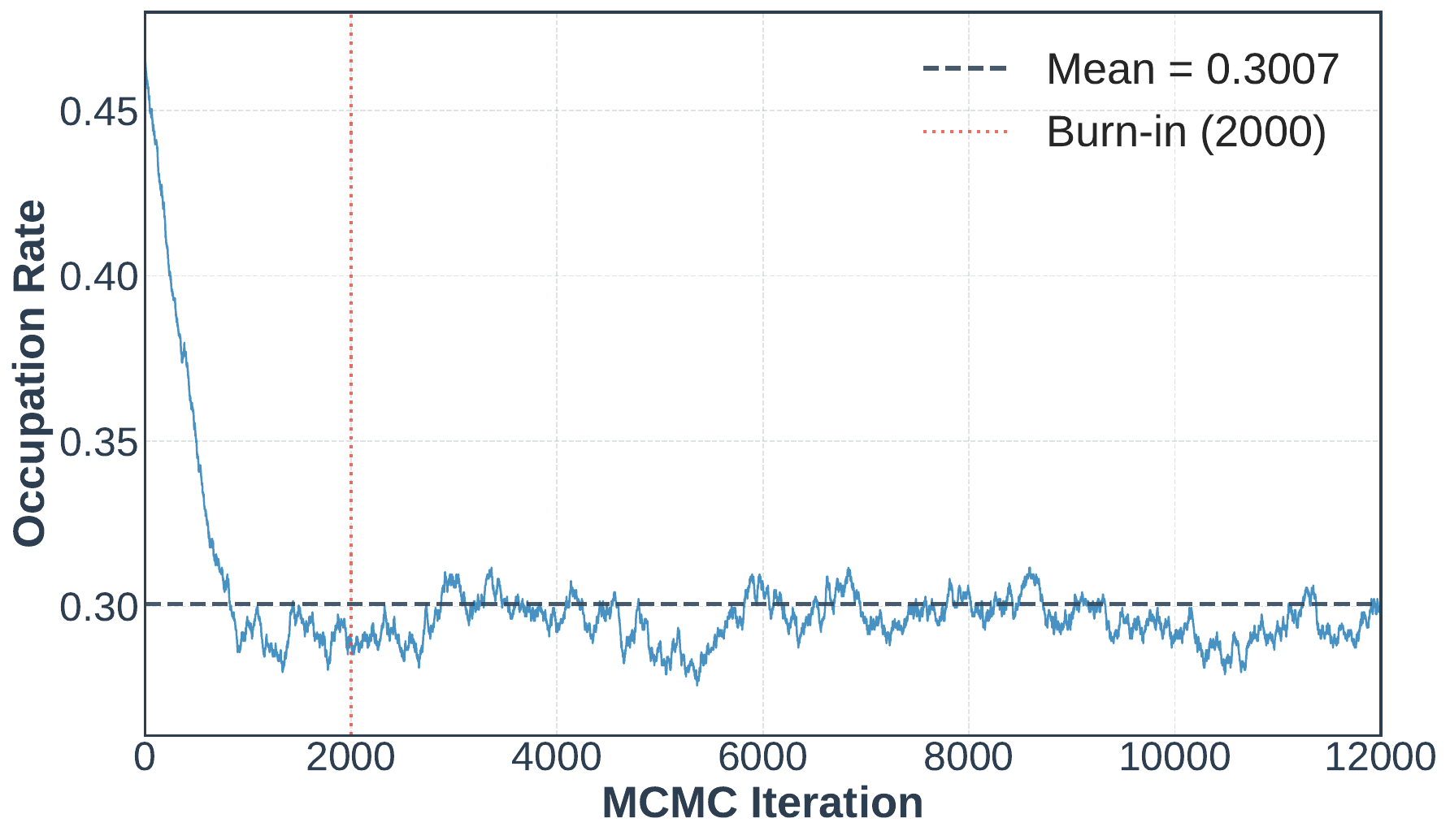}
\label{fig: occ_pwwa_pam50}
\end{subfigure}
\begin{subfigure}{.48\textwidth}
\centering
\includegraphics[width=0.9\linewidth]{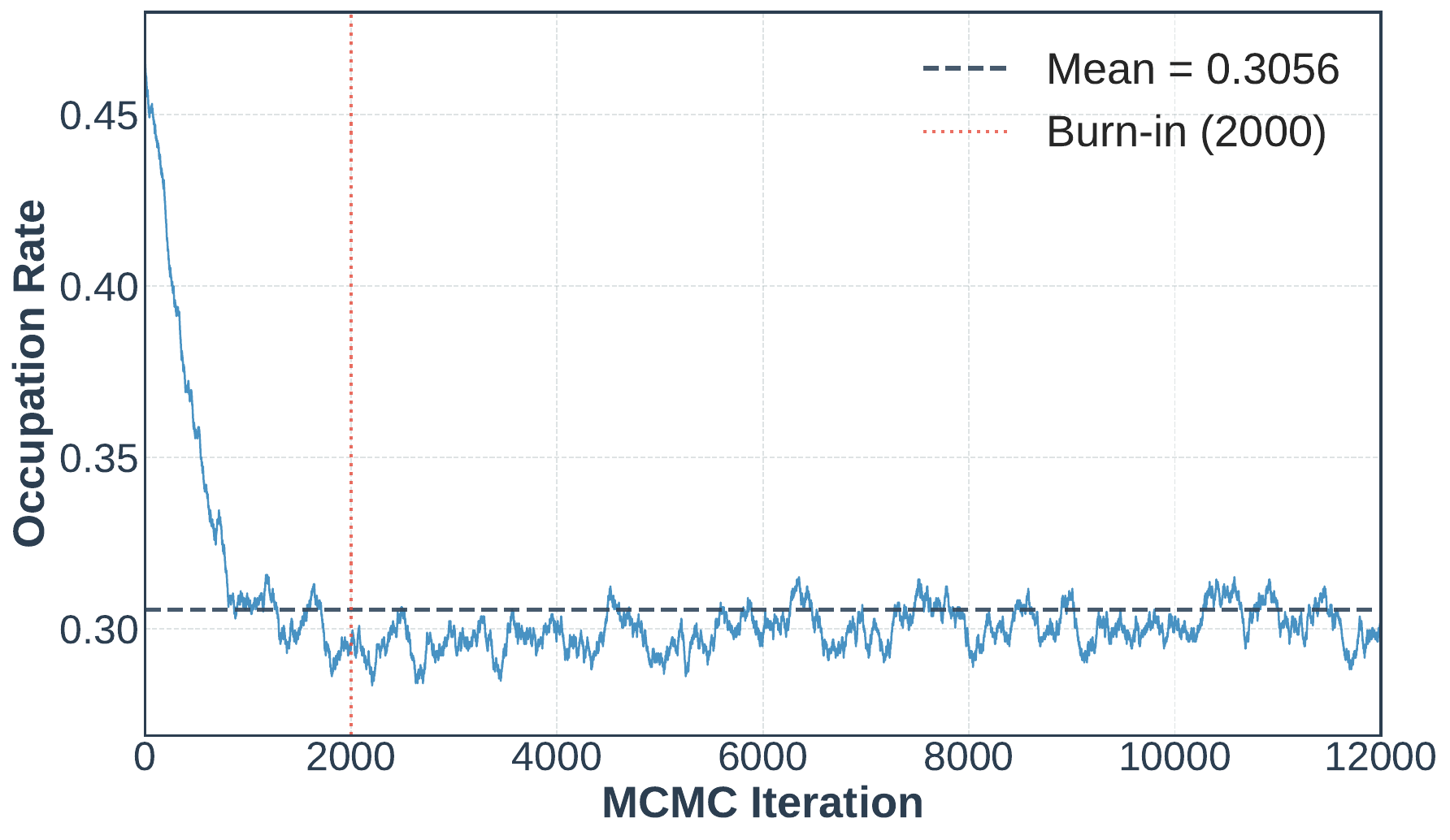}
\label{fig: occ_pbd_pam50}
\end{subfigure}
\caption{Trace plots of the posterior simulations for the PAM50 breast cancer dataset obtained using four Bayesian posterior simulation algorithms. 
The panels correspond to copula-WWA (upper left), copula-BD (upper right), probit-WWA (lower left), and probit-BD (lower right), respectively.}
\label{fig: occ_pam50}
\end{figure}

To assess the convergence and mixing behavior of the MCMC samplers, Figure~\ref{fig: occ_pam50} presents the 
trace-plots of the `occupation rate', 
the latter quantity being defined as the proportion of present edges among all possible edges in the sampled graphs. 
The trajectories of the occupation rates indicate satisfactory mixing and effective exploration of the posterior graph space for all four methods.

We next examine the biological interpretation of the inferred networks. 
A prominent feature is that several key genes from the PAM$50$ dataset, including ERBB2, ESR1, FOXC1, NAT1, and SFRP1, are connected to multiple breast cancer subtypes. 
Overall, the inferred networks are highly consistent with established biological knowledge. In particular, ESR1 and ERBB2, which encode the estrogen receptor (ER) and HER2 receptor, respectively, are consistently identified as hub genes linking multiple molecular subtypes, thus our fidings reflect the central roles of these genes in breast cancer classification.
Likewise, FOXC1 is repeatedly associated with the basal subtype, in agreement with its established role as a marker of basal-like breast cancer and poor clinical outcome~\cite{10.1158/1055-9965.EPI-13-1017,10.1158/0008-5472.CAN-09-4120,jensen2015diagnosis}. In agreement with previous molecular profiling studies, NAT1 is associated with luminal subtypes~\cite{doi:10.1073/pnas.96.16.9212,doi:10.1073/pnas.191367098}, while SFRP1 exhibits subtype-dependent associations reported in both luminal and basal-like breast cancers~\cite{huth2014bdnf}. Furthermore, an BCL2--LumA association is recovered by three of the four algorithms, and this result is consistent with the reported enrichment of BCL2 expression in luminal A tumors and its favorable prognostic value~\cite{eom2016bcl2}.

The four algorithms produced highly consistent network structures, 
recovering between $84$ and $104$ subtype--gene edges, with $187$ edges shared across all methods. 
The copula-based methods (copula-WWA and copula-BD) and the probit-based methods (probit-WWA and probit-BD) each yielded nearly identical sets of edges, 
suggesting that the inferred biological relationships are robust to the choice of computational algorithm. 
Overall, the proposed methods successfully recover biologically meaningful dependency structures from mixed continuous and discrete data, 
demonstrating their effectiveness for Bayesian structure learning in MGMs.

\begin{figure}[!htbp]
\begin{subfigure}{.48\textwidth}
\centering
\includegraphics[width=0.9\linewidth]{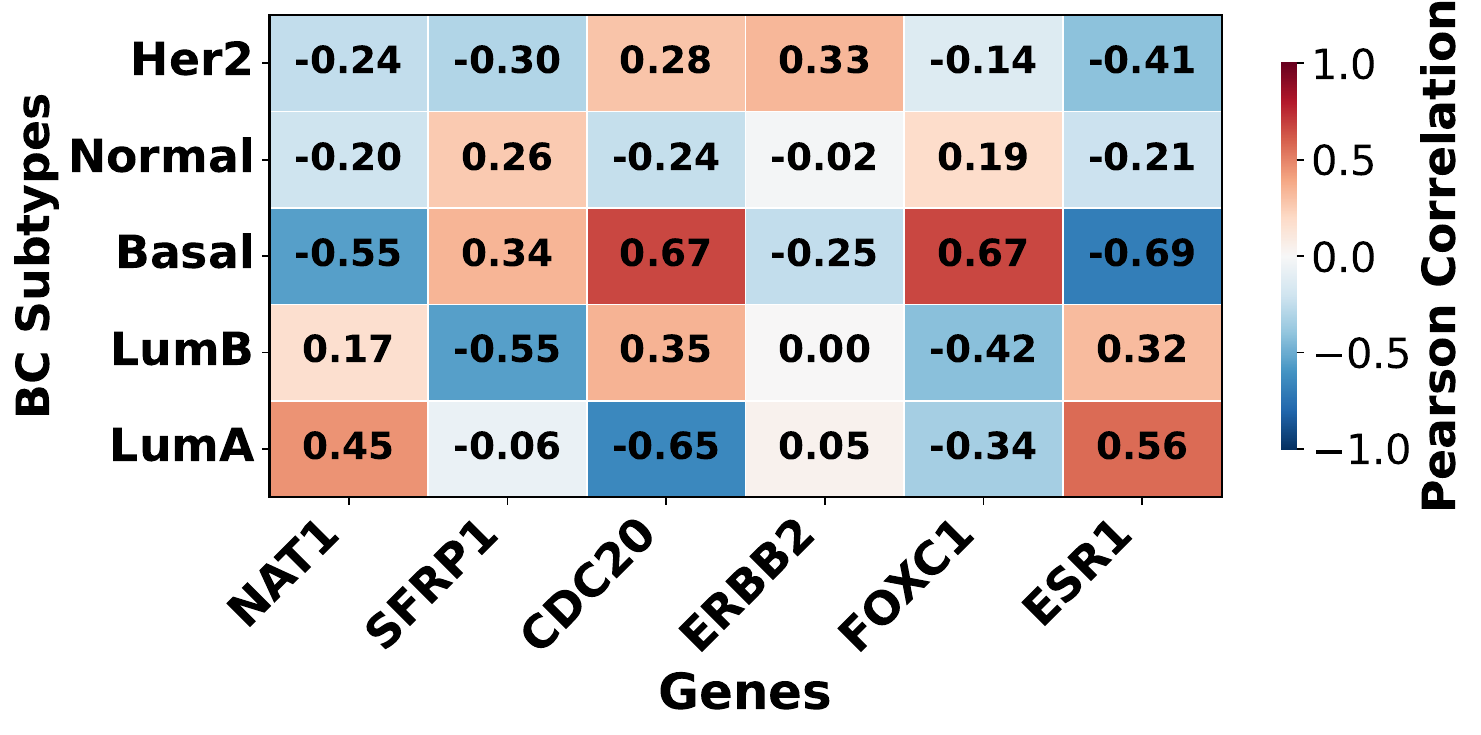}
\caption{Copula-WWA}
\label{fig: corr_cwwa_pam50}
\end{subfigure}
\begin{subfigure}{.48\textwidth}
\centering
\includegraphics[width=0.9\linewidth]{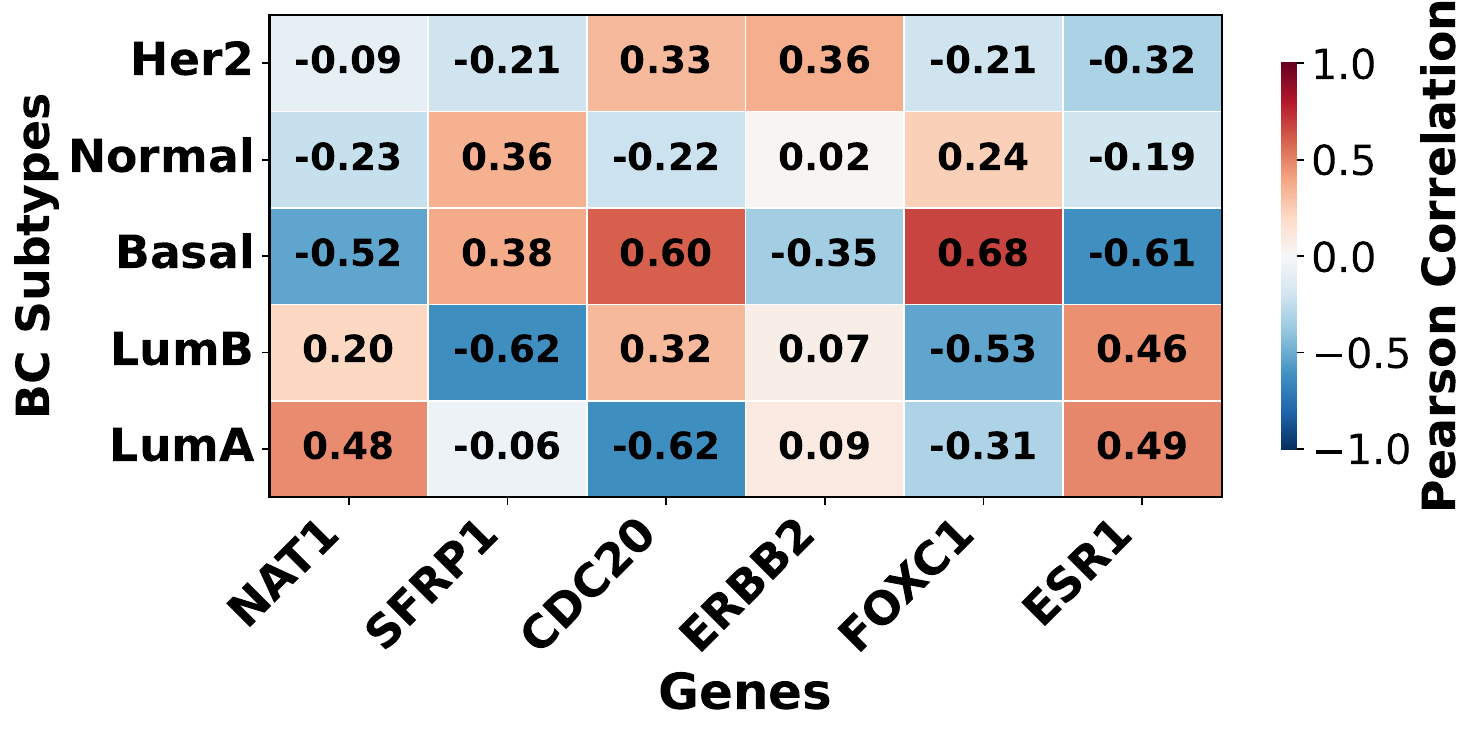}
\caption{Copula-BD}
\label{fig: corr_cbd_pam50}
\end{subfigure}\\
\begin{subfigure}{.48\textwidth}
\centering
\includegraphics[width=0.9\linewidth]{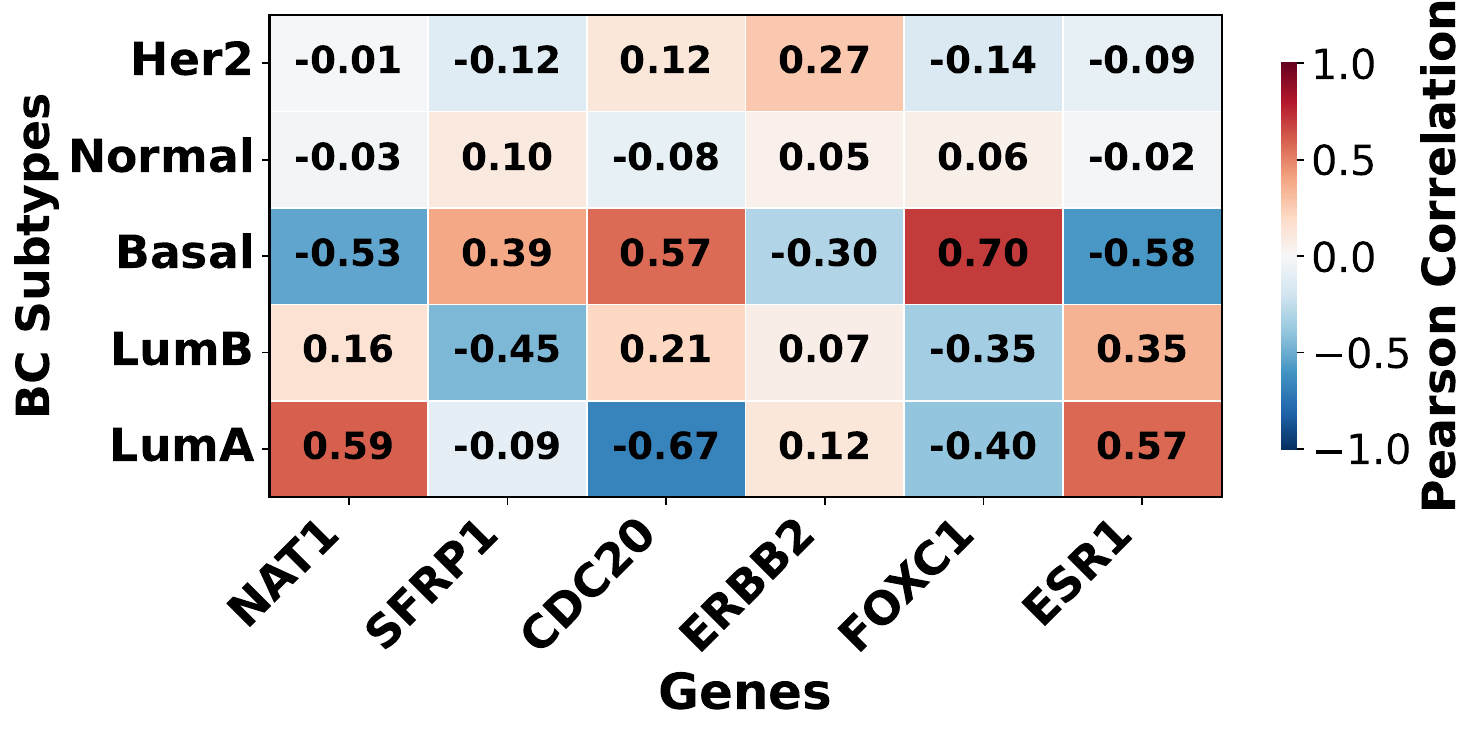}
\caption{Probit-WWA}
\label{fig: corr_pwwa_pam50}
\end{subfigure}
\begin{subfigure}{.48\textwidth}
\centering
\includegraphics[width=0.9\linewidth]{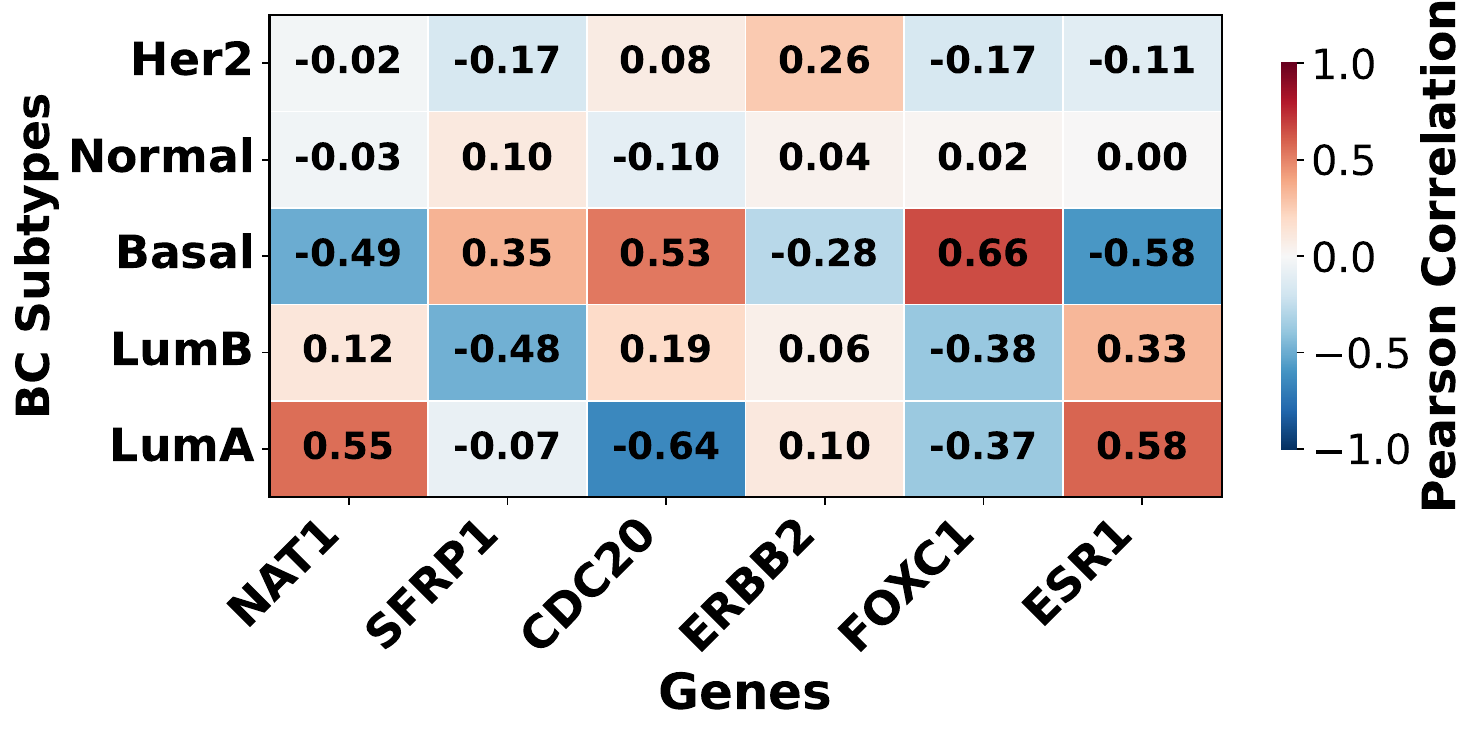}
\caption{Probit-BD}
\label{fig: corr_pbd_pam50}
\end{subfigure}
\caption{Pearson correlation matrices between breast cancer subtypes and selected gene expressions identified by copula-WWA (upper left), copula-BD (upper right), probit-WWA (lower left), and probit-BD (lower right).}
\label{fig: corr_pam50}
\end{figure}

As illustrated in the Pearson correlation matrices in Figure~\ref{fig: corr_pam50}, 
the associations between PAM$50$ gene expressions and breast cancer subtypes derived from the estimated precision matrices of all four methods exhibit high consistency and showcase biologically meaningful patterns.
Across all methods, ESR1 is strongly negatively correlated with the basal subtype and positively correlated with LumA, in agreement with its established role as a marker of estrogen receptor-positive breast cancer. 
Similarly, FOXC1 and CDC20 exhibit positive associations with the basal subtype, whereas CDC20 is negatively associated with LumA, in agreement with their reported roles in basal-like and proliferative breast cancers. 
ERBB2 shows positive correlations with the HER2 subtype, while NAT1 is positively associated with LumA and negatively associated with the basal subtype. 
SFRP1 displays negative correlations with LumB, and this is consistent with SFRP1's reported tumor suppressor function.
Although the overall correlation structures are highly concordant across methods, modest quantitative differences are observed. 
In particular, copula-WWA (Figure~\ref{fig: corr_cwwa_pam50}) tends to recover stronger correlations for proliferation related markers, 
whereas probit-WWA (Figure~\ref{fig: corr_pwwa_pam50}) produces stronger associations for several luminal markers. 
Overall, the four methods recover biologically interpretable dependency structures 
that are consistent with established molecular mechanisms of breast cancer, 
supporting the biological validity of the estimated MGMs.

\begin{figure}[!th]
\centering
\includegraphics[width=0.98\linewidth]{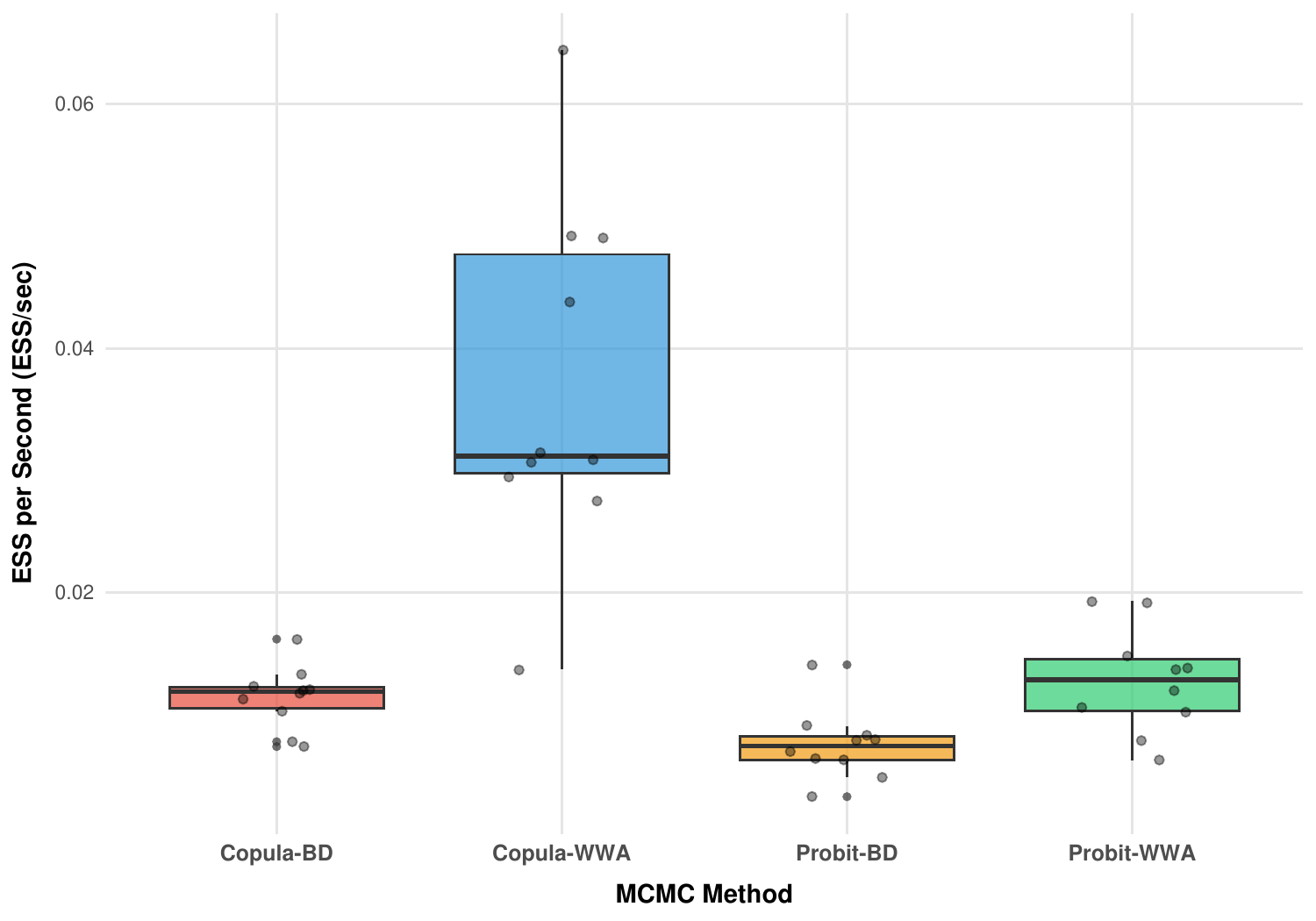}
\caption{
Boxplots of the effective sample size per second (ESS/s) obtained from our four MCMC algorithms.
Higher values indicate greater sampling efficiency relative to computational cost.
}
\label{fig:ess_per_second}
\end{figure}
Figure~\ref{fig:ess_per_second} reports the effective sample size per unit computational time (ESS/sec), 
which jointly quantifies sampling efficiency and computational cost. 
Both WWA-based algorithms consistently outperform the corresponding BD-based methods, 
demonstrating that the proposed M-WWA framework provides a more efficient mechanism for posterior exploration. 
Notably, the efficiency gain is considerably larger for the copula-based formulation with copula-WWA showing a substantially greater improvement over copula-BD contrasted to the case of probit-WWA over probit-BD. 
This suggests that the benefits of the M-WWA framework are robust across both latent variable formulations, while being particularly pronounced under the copula representation.
Further results from the simulation analysis of PAM$50$ are provided in Appendix~\ref{app: additional_results_pam50}.

\section{Discussion}
\label{sec: diss}
In this work, we introduced the M-WWA methodology, a unified framework for Bayesian posterior simulation in MGMs.
By extending WWA of \cite{van2022g} to latent variable formulations of MGMs, we developed two concrete instantiations, copula-WWA and probit-WWA, tailored to copula-MGMs and probit MGMs, respectively.
These methods address key computational challenges that arise in Bayesian structure learning for heterogeneous data. 

The numerical experiments with simulated data demonstrate that both copula-WWA and probit-WWA consistently 
outperform their BD-based counterparts in terms of structure recovery accuracy, 
particularly as the graph dimension increases.
The gains are most pronounced in moderate-to-small sample regimes, where efficient exploration of the graph space is critical. 
These results suggest that the combination of 
latent variable modeling with the informed proposal mechanism, locally-balanced graph exploration, 
and delayed-acceptance strategy provides an effective framework for posterior simulation in high-dimensional MGMs. 
Compared with conventional methods that explore the graph space without utilising posterior information, 
the integrated M-WWA framework can achieve improved mixing and computational efficiency
by facilitating more targeted exploration of high-probability graph configurations.

The real-data application to the PAM$50$ breast cancer dataset further illustrates the practical relevance of the proposed methods.
The inferred MGMs capture interpretable dependencies between gene expression levels and molecular subtypes, while the trace-plots indicate fast convergence and mixing. 
In addition, the computational comparison demonstrates that the WWA-based approaches achieve higher ESS/sec than their corresponding BD-based counterparts, with copula-WWA outperforming copula-BD, and probit-WWA outperforming probit-BD. 
Importantly, the ability to jointly model continuous gene expression measurements and discrete subtype indicators within a coherent Bayesian framework highlights the flexibility of the proposed approach for integrative biological data analysis and its use in a multitude of other fields where structure learning for heterogeneous variable is of interest.


It terms of further contributions, scaling up M-WWA to tackle very high-dimensional settings represents a crucial avenue for future research. 
Optimizing current computational bottlenecks, particularly within graph-space exploration, will be essential to enable efficient posterior simulation for networks with thousands of nodes while maintaining practical execution costs.
A more exhaustive analysis of the mixing properties of M-WWA, particularly in comparison with continuous-time popular alternative methods, represents an important avenue for future investigation.

Overall, M-WWA provides a principled, scalable approach to Bayesian structure learning for MGMs, bridging the gap between methodological flexibility and computational efficiency.
We expect that the proposed framework will serve as a foundation for further developments in Bayesian inference for complex heterogeneous data.

\section{Acknowledgements}
Erdong Guo thanks Kayvan Sadeghi, Louis Sharrock, Sam Livingstone, Reza Mohammadi, and Kengo Kamatani for helpful discussions. 
Erdong Guo also thanks Nanwei Wang for sharing the PAM$50$ dataset. 
The authors acknowledge the use of the Myriad High-Performance Computing Facility (Myriad@UCL), 
and the associated support services, in the completion of this work.

\bibliography{reference.bib}
\clearpage

\appendix


\section{Details for Simulation Studies}
\label{app: simu_details}
Synthetic datasets are generated given graph  structures
with dimension $p$ and varying sample sizes $n$. 
Each dataset comprised mixed variable types: $50\%$ continuous (Gaussian), $20\%$ binary,
and $30\%$ ordinal variables with $5$ levels. 
The underlying precision matrix was constructed with edge probability $p$ under the given topology.
Specifically, conditional on the graph $G$, 
we sampled the precision matrix $K$ from the G-Wishart distribution $W_G(3,I_p)$.
For each of the four competing algorithms, experiments were replicated awith independent random seeds, yielding unique experimental configurations per method. 
MCMC computations utilized $10{,}000$ burn-in iterations and $40{,}000$ sampling iterations, with M-WWA methods configured for $n_e=2$ updates per iteration.

\subsection{Graph Structures Used in Simulation Studies}
To evaluate performance under diverse conditional dependence patterns, we consider several canonical graph topologies commonly used in the graphical modeling literature. Let $p$ denote the number of nodes and $K = (K_{i,j})$ the corresponding precision matrix.

\paragraph{Cycle Graph.}
Nodes are arranged on a ring so that each node is connected primarily to its immediate neighbors. Specifically, diagonal entries for the true precision matrix satisfy $K_{i,i}=1$ for all $i$. First-order neighbors along the cycle have edge weights $K_{i,i-1}=K_{i-1,i}=0.5$, while the wrap-around edge between nodes $1$ and $p$ is assigned $K_{1,p}=K_{p,1}=0.4$. All remaining off-diagonal entries are set to zero.

\paragraph{Star Graph.}
This topology contains a single hub node connected to all other nodes. The true precision matrix satisfies $K_{i,i}=1$ for all $i$, while for $i \neq 1$ we set $K_{1,i}=K_{i,1}=0.1$. No additional edges are present, so $K_{i,j}=0$ for all other off-diagonal pairs.

\paragraph{AR(1) Graph.}
Dependence decays exponentially with the distance between node indices. In particular, the covariance matrix $\Sigma = (\Sigma_{i,j})$ is specified by
\[
\Sigma_{ij} = 0.7^{|i-j|},
\]
thus $\Sigma$ induces a banded dependence structure characteristic of a first-order autoregressive process.

\paragraph{AR(2) graph.}
This structure extends local dependence to second-order neighbors. Diagonal elements satisfy $K_{i,i}=1$. First-order neighbors have weights  $K_{i,i-1}=K_{i-1,i}=0.5$, and second-order neighbors have weights $K_{i,i-2}=K_{i-2,i}=0.25$. All other off-diagonal elements are zero.

\paragraph{Random Graph.}
Edges are generated independently for each unordered node pair with inclusion probability $2/(p-1)$. Conditionally on the resulting graph $G$, the precision matrix is sampled from a G-Wishart distribution:
\[
K \sim \mathcal{W}_G(3, I_p),
\]
thus ensuring positive definiteness while respecting the graph constraints.

\paragraph{Clustered Graph.}
Nodes are partitioned into $\mathrm{cl} = \max\{2, \lfloor p/20 \rfloor\}$ disjoint clusters. Within each cluster, edges are generated according to the same random graph mechanism described just above, while no edges are allowed between clusters. The precision matrix is then drawn from:
\[
K \sim \mathcal{W}_G(3, I_p).
\]

\paragraph{Scale-free graph.}
A scale-free topology is constructed using the Barabási–Albert preferential attachment mechanism, which produces a network with hub nodes and a heavy-tailed degree distribution. The resulting graph contains approximately $(p-1)$ edges. As with the previous stochastic graph settings, the precision matrix is sampled from
\[
K \sim \mathcal{W}_G(3, I_p).
\]

\section{Additional Results for the PAM$50$ Dataset}
\label{app: additional_results_pam50}
\begin{figure}[!th]
\centering
\includegraphics[width=0.98\linewidth]{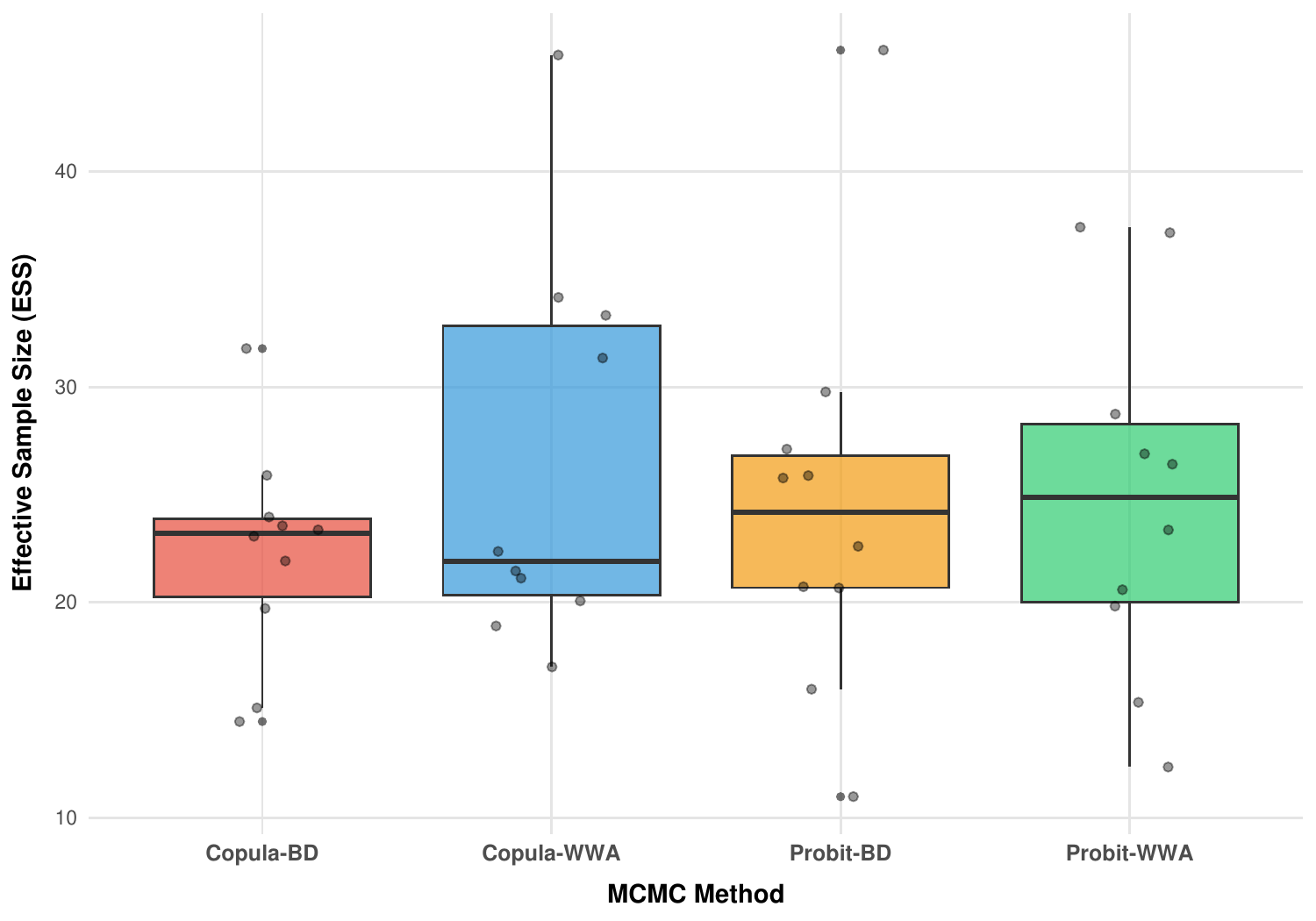}
\caption{
Boxplots of the ESS for the four MCMC algorithms (copula-WWA, probit-WWA, copula-BD, and probit-BD)
under the simulation settings described in Section~\ref{sec: real}.
}
\end{figure}

\begin{figure}[!th]
\centering
\includegraphics[width=0.98\linewidth]{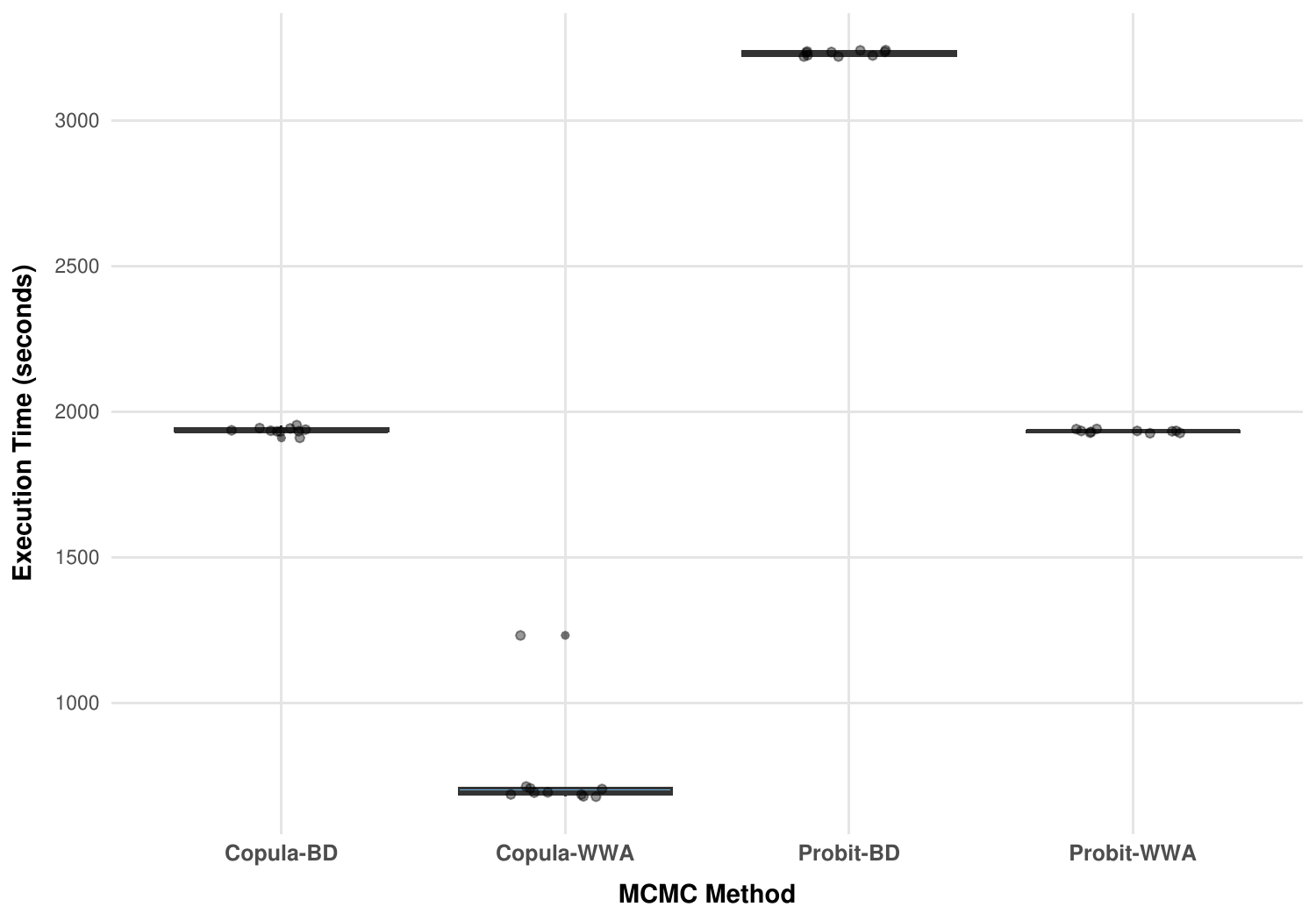}
\caption{
Boxplots of the total execution time for the four MCMC algorithms (copula-WWA, probit-WWA, copula-BD, and probit-BD)
under the simulation settings described in Section~\ref{sec: real}
}
\end{figure}

\begin{figure}[!ht]
\begin{subfigure}{.5\textwidth}
\centering
\includegraphics[width=1\linewidth]{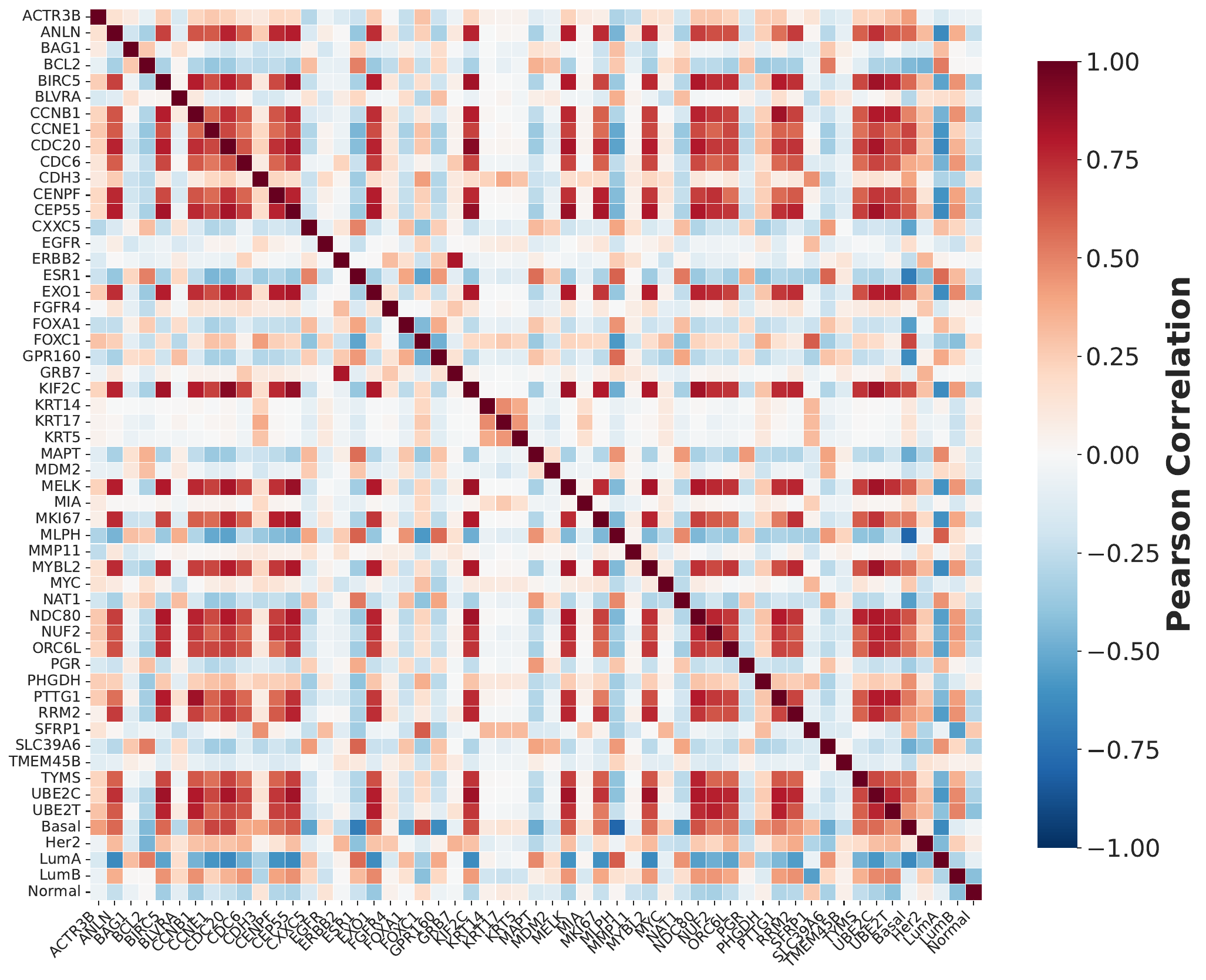}
\caption{Copula-WWA}
\label{fig: corr_full_cwwa_pam50}
\end{subfigure}
\begin{subfigure}{.5\textwidth}
\centering
\includegraphics[width=1\linewidth]{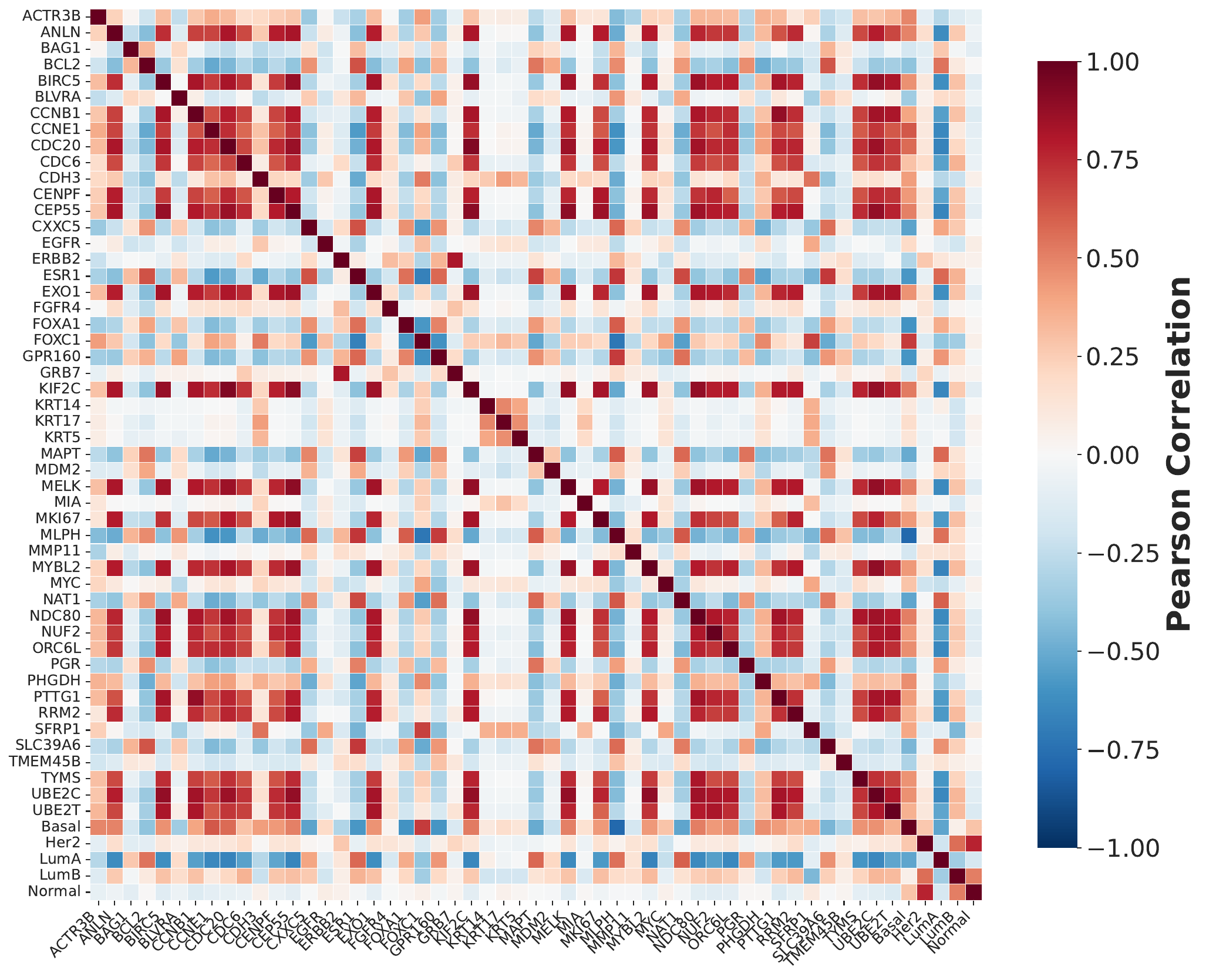}
\caption{Probit-WWA}
\label{fig: corr_full_pwwa_pam50}
\end{subfigure}\\
\begin{subfigure}{.5\textwidth}
\centering
\includegraphics[width=1\linewidth]{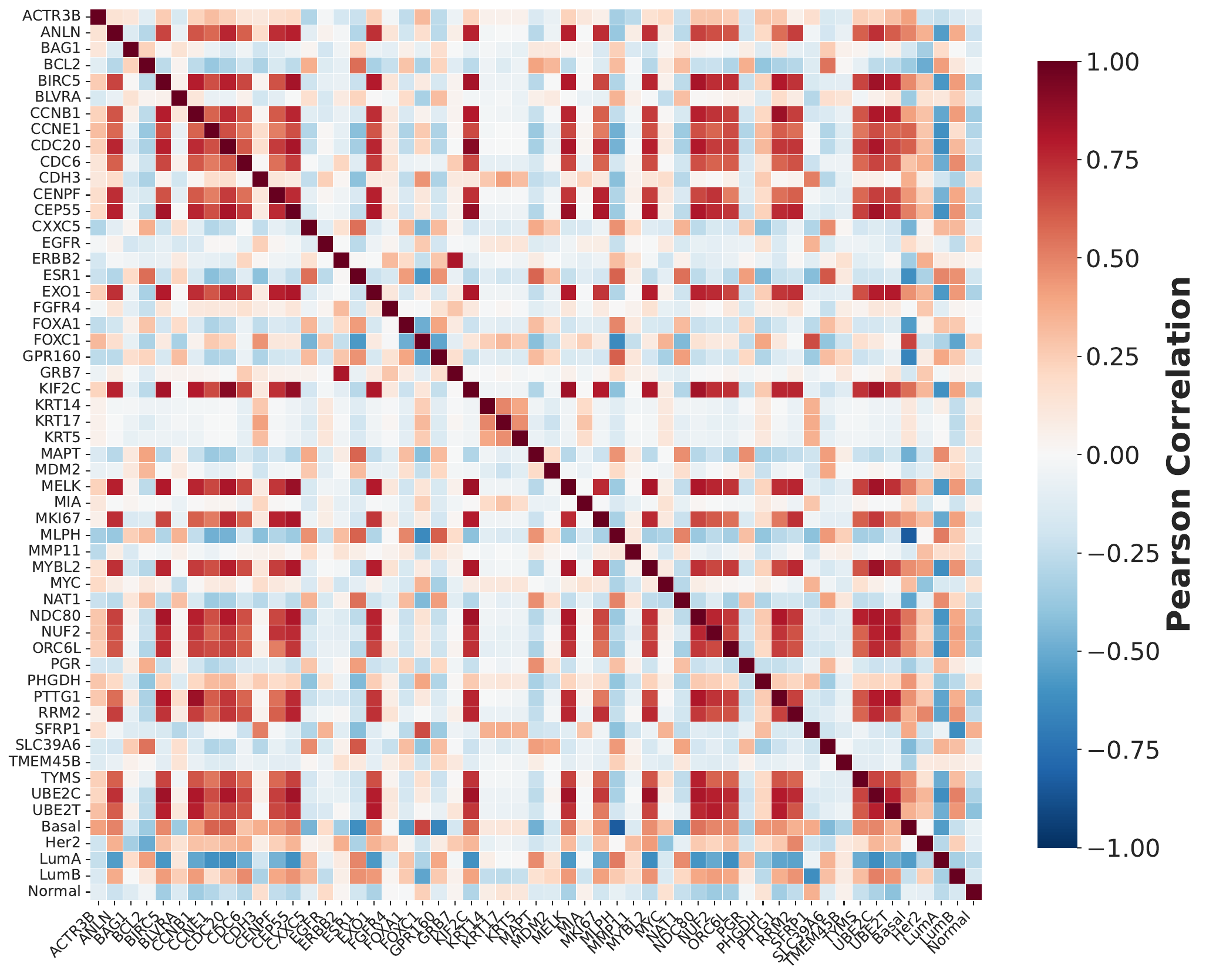}
\caption{Copula-BD}
\label{fig: corr_full_cbd_pam50}
\end{subfigure}
\begin{subfigure}{.5\textwidth}
\centering
\includegraphics[width=1\linewidth]{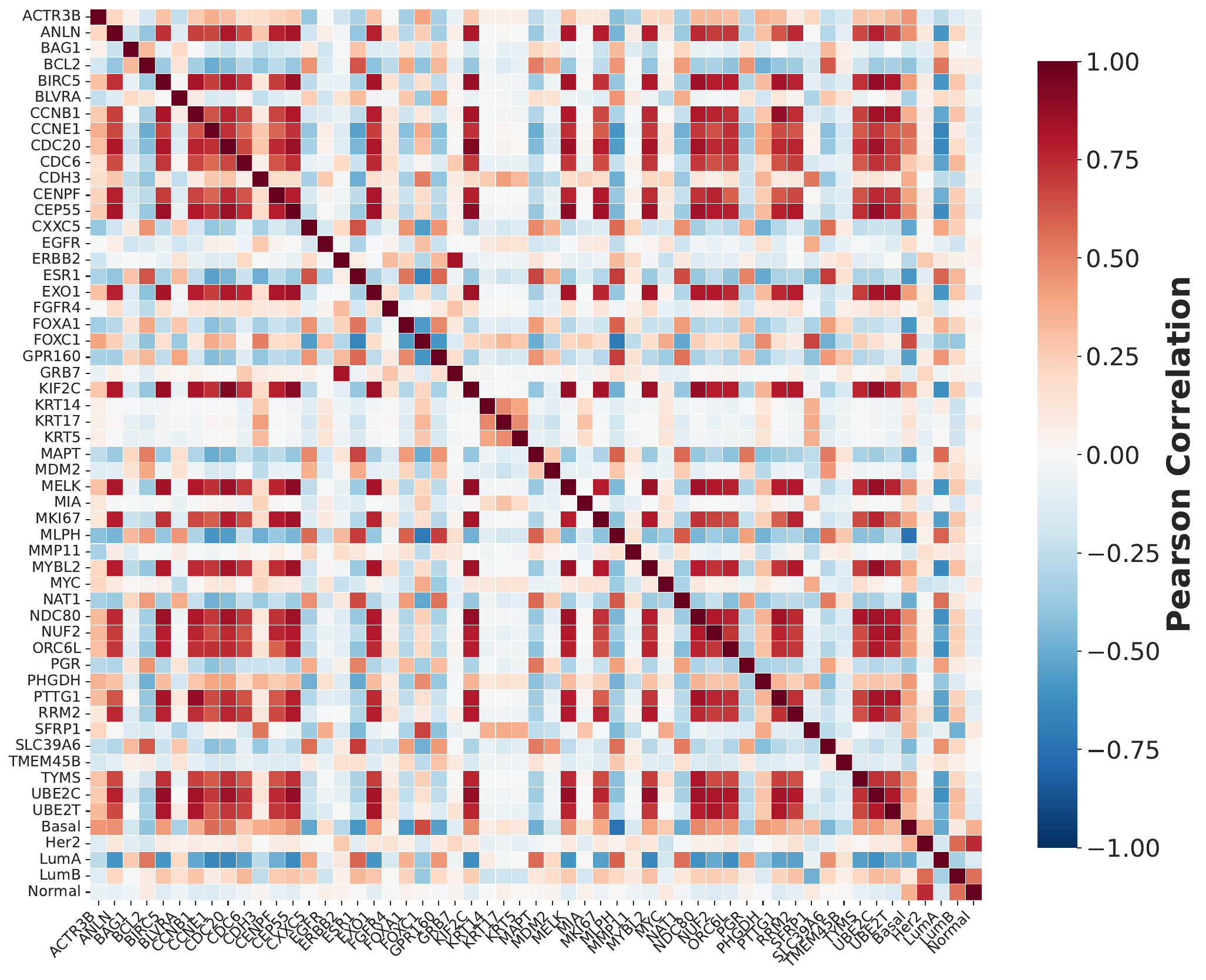}
\caption{Probit-BD}
\label{fig: corr_full_pbd_pam50}
\end{subfigure}
\caption{Pearson correlation matrices estimated by copula-WWA, copula-BD, probit-WWA, and probit-BD algorithms for the PAM$50$ dataset.}
\label{fig: corr_full_pam50}
\end{figure}

\end{document}